\documentstyle[12pt,graphicx]{article}                
\begin{document}
%This is dvips(k) 5.86 Cobegin{document}
\baselineskip 18pt
%t
\def\today{\ifcase\month\or
 January\or February\or March\or April\or May\or June\or
 July\or August\or September\or October\or November\or Decembe\fi
 \space\number\day, \number\year}
\def\thebibliography#1{\section*{References\markboth
 {References}{References}}\list
 {[\arabic{enumi}]}{\settowidth\labelwidth{[#1]}
 \leftmargin\labelwidth
 \advance\leftmargin\labelsep
 \usecounter{enumi}}
 \def\newblock{\hskip .11em plus .33em minus .07em}
 \sloppy
 \sfcode`\.=1000\relax}
\let\endthebibliography=\endlist
\def\lsim{\ ^<\llap{$_\sim$}\ }
\def\gsim{\ ^>\llap{$_\sim$}\ }
\def\r2{\sqrt 2}
\def\beq{\begin{equation}}
\def\eeq{\end{equation}}
\def\beqn{\begin{eqnarray}}
\def\eeqn{\end{eqnarray}}

\begin{titlepage}

\begin{center}
{\large {\bf Effective Lagrangian for the $\chi^{+}_j \chi^{-}_kH^{0}_l$ interaction in the minimal supersymmetric standard model and neutral Higgs decays}}\\
\vskip 0.5 true cm
\vspace{2cm}
 Tarek Ibrahim  
\vskip 0.5 true cm
\end{center}

\noindent
\begin {center}
{Department of Physics, Northeastern University, Boston, MA 02115-5000, USA }\\
{and}\\
{Department of  Physics, Faculty of Science,
University of Alexandria, Alexandria, Egypt
\footnote{Permanent address.}}\\
\end{center} 
\vskip 1.0 true cm
\centerline{\bf Abstract}
\medskip
We extend previous analyses of the supersymmetric loop correction to the  neutral Higgs couplings to include the coupling $\chi^{+}_j \chi^{-}_kH^{0}_l$. The analysis completes the previous analyses where similar corrections were computed for the $\bar{\tau} \tau H^{0}_l$, $\bar{b} b H^{0}_l$, $\bar{c} c H^{0}_l$ and for $\bar{t} t H^{0}_l$ couplings within the minimal supersymmetric standard model. The effective one loop Lagrangian is then applied to the computation of the neutral Higgs decays. 
The sizes of the supersymmetric loop corrections of the neutral Higgs decay widths
into $\chi^{+}_i \chi^{-}_j$ ($i=1,2$; $j=1,2$) are investigated and the supersymmetric
loop correction is found to be in the range of $7\sim15\%$ in 
significant   regions  of the parameter space. By including the loop corrections
of the other decay channels $\bar{b} b$, $\bar{t} t$, $\bar{\tau} \tau$, $\bar{c} c$, 
and $\chi^0_i \chi^0_j$ ($i=1-4$; $j=1-4$), the corrections   to  branching ratios for
$H^{0}_l\rightarrow  \chi^{+}_i \chi^{-}_j$ can reach as high as $40\%$. 
 The effects of CP phases on the branching ratio are also investigated.
\end{titlepage}
\section{INTRODUCTION}

The neutral Higgs couplings to different fields are of great current interest as they enter in a variety of phenomena which are testable in low energy processes \cite{carena1}.
It is known that supersymmetric corrections can affect the neutral Higgs boson decays into $b\bar{b}$, $\tau \bar{\tau}$ and $c\bar{c}$. The decay properties of the lightest Higgs boson in MSSM would be different from those of the Standard Model Higgs boson when these corrections are taken into consideration. Specifically the ratio of the branching ratios to $b\bar{b}$ and $\tau \bar{\tau}$ of the Higgs boson is an important piece of evidence that might distinguish between the lightest MSSM Higgs boson and the Standard Model one at colliders.
In MSSM there are also other modes for neutral Higgs decays that do not exist in Standard Model such as charginos and neutralinos.

In this paper we compute the one loop corrected effective
 Lagrangian for the neutral Higgs and chargino couplings. We then analyze the effects of the loop corrections to the neutral Higgs decays 
$H^0_l \rightarrow \chi^+_j\chi^-_k$. In the analysis we also include the effect of CP phases arising from the soft SUSY breaking parameters. It is well known that large CP phases can be made compatible \cite{edm1,edma,edmb} with experimental constraints on the electric dipole moments (edms) of the electron \cite{edm2}, of the neutron \cite{edm3}, and of the $Hg^{199}$ \cite{edm4}.
Further, if the phases are large they could affect the Higgs
sector physics. It is well known that one loop contributions
to the Higgs masses from the stop, sbottom, the
chargino and neutralino sectors can lift the lightest
Higgs mass above $M_Z$. The inclusion of the CP violating
phases brings mixings between the CP even and the CP odd
Higgs \cite{edm5,edm5a,edm5b,demir1,we4,we5}.
The CP violating phases modifies the physics of dark matter \cite{edm6}, and of other
 phenomena \cite{edm7}.  (For a  review see  Ref.\cite{cpreview}.)

The current analysis of $\Delta {\cal{L}}_{H^{0}\chi^{+}\chi^{-}}$ and 
neutral Higgs decay into charginos is based on the effective Lagrangian
method where the couplings of the electroweak eigen states $H^{1}_1$ and 
 $H^{2}_2$ with charginos are radiatively corrected using the zero external
momentum approximation. The same technique has been used in calculating the 
effective Lagrangian and decays of $H^{0}_l$ into quarks and leptons \cite{carena1,babu,we1}.
It has been used also in the analysis of the effective Lagrangian of charged Higgs with quarks \cite{carena1,two} and their decays into ${\bar{t}}b$ and $\nu_{\tau} \tau$
\cite{we2}
and into chargino $+$ neutralino \cite{we3}.
The neutral Higgs decays into charginos have been investigated before in the
CP conserving case \cite{eberl,ren}. In these analyses, the wave function
renormalization and the counter terms for the mass matrix elements are 
calculated beside the vertex corrections of the mass eigen states $h^{0}$, 
$H^{0}$ and $A^{0}$ with charginos. In the effective Lagrangian technique 
with zero external momentum approximation, the radiative corrections
of the processes considered here originate only from the vertex contributions.
Thus our analysis of the neutral Higgs decays into charginos is a partial one.
However,  as mentioned before the above analyses were carried out in the CP 
conserving scenario. As far as we know, the analysis for the neutral Higgs 
decays into charginos, with one loop corrections, in the CP violating case where the neutral Higgs sector
is modified in couplings, spectrum and mixings, does not exist.
We evaluate the radiative corrections to the Higgs boson masses
and mixngs by using the effective potential approximation.
We include the corrections from the top and bottom quarks and squarks 
\cite{demir1}, from the chargino, the W and the charged 
Higgs sector \cite{we4} and from the neutralino, Z boson, and 
the neutral Higgs bosons \cite{we5}.
It is important to notice that the corrections to the Higgs effective
potential from the different sectors mentioned above are all one-loop corrections.
The corrections of the interaction $\Delta {\cal{L}}_{H^{0}\chi^{+}\chi^{-}}$
to be considered in this work are all one-loop level ones. So the analysis
presented here is a consistent one loop study. 

The outline of the rest of the paper is as follows: In Sec. 2 we compute the effective Lagrangian for the $\chi^{+}_j \chi^{-}_kH^{0}_l$ interaction. In Sec. 3 we give an analysis of the decay widths of the neutral Higgs bosons into charginos using the effective Lagrangian. In Sec. 4 we give a numerical analysis of the size of the loop effects on the partial decay width and on the  branching ratios. Conclusions are given in Sec. 5.

\section{LOOP CORRECTIONS TO NEUTRAL HIGGS COUPLINGS}

The tree-level Lagrangian for $\chi^{+}_j \chi^{-}_kH^{0}$ interaction is
\beq
{\cal{L}}=\phi_{jk}\overline{\chi_{j}^{+}} P_R \chi^{+}_k H^{1}_1+
\psi_{jk}\overline{\chi_j^{+}} P_R \chi^{+}_k H^{2}_2+H.c.,
\eeq
where $H^1_1$ and $H^2_2$ are the neutral states of the two Higgs isodoublets in the minimal supersymmetric standard model (MSSM), i.e.,
\beqn
(H_1)= \left(\matrix{H_1^1\cr
 H_1^2}\right),~~
(H_2)= \left(\matrix{H_2^1\cr
             H_2^2}\right)
\eeqn
and the couplings $\phi_{jk}$ and $\psi_{jk}$ are given by
\beq
\phi_{jk}=-g U_{k2} V_{j1},~~ \psi_{jk}=-g U_{k1} V_{j2}
\eeq
where U and V diagonalize the chargino mass matrix so that
\beq
U^{*} M_{\chi^{+}}V^{-1}=diag(m_{\chi^{+}_1},m_{\chi^{+}_2})
\eeq
The loop corrections produce shifts in the couplings of Eq. (1) and the effective Lagrangian with loop corrected couplings is given by
\beqn
{\cal{L}}_{eff}=(\phi_{jk}+\delta \phi_{jk})\overline{\chi_j^{+}} P_R \chi^{+}_k H^{1}_1+ \Delta \phi_{jk} \overline{\chi_j^{+}} P_L \chi^{+}_k H^{2}_2+\nonumber\\
~~(\psi_{jk}+\delta \psi_{jk})\overline{\chi_j^{+}} P_R \chi^{+}_k H^{2}_2+ \Delta \psi_{jk} \overline{\chi_j^{+}} P_L \chi^{+}_k H^{1}_1
+H.c.
\eeqn
In this work we calculate the loop correction to the $\chi^{+}_j \chi^{-}_kH^{0}_l$ using the zero external momentum approximation.

\subsection{Loop analysis of $\delta \phi_{jk}$ and  $\Delta \psi_{jk}$}
Contributions to $\delta \phi_{jk}$ and  $\Delta \psi_{jk}$ arise from the thirteen loop diagram of Fig. 1. We note that the contribution from diagrams which have $H^+ W^+ H^0$and $H^0 Z^0 H^0$ vertices do not contribute in the
effective Lagrangian with 
 zero external momentum approximation since these vertices are proportional to the external momentum. We discuss now in detail the contribution of each of these diagrams in Fig. 1.  We begin with the loop diagram of Fig. 1i(a) which contributes to $\delta \phi_{jk}$ and  $\Delta \psi_{jk}$.

We calculate the corrections of the amplitude from Fig. 1i(a) 
\beq
\delta M =i \delta \phi_{jk} \bar{u}_jP_R v_k +i \Delta \psi_{jk} \bar{u}_jP_L v_k
\eeq 
The idea is to extract, from the amplitude correction, the expressions for
$\delta \phi_{jk}$ and  $\Delta \psi_{jk}$ from those parts that are
proportional to $\bar{u}_jP_R v_k$ and $\bar{u}_jP_L v_k$ respectively.
For this purpose we need $\tilde{b} \tilde{b} H^1_1$ interaction which
is given by
\beq
{\cal{L}}_{\tilde{b}\tilde{b}H^{1}_1}=H_{il} \tilde{b}^{*}_i \tilde{b}^{*}_l H^{1}_1 +H.c. 
\eeq 
where $H_{il}$ is given by
\beqn
H_{il}=-\frac{gM_Z}{\sqrt 2 \cos\theta_W}
((-\frac{1}{2}+\frac{1}{3}\sin^2\theta_W)D_{b1i}^{*}D_{b1l}
-\frac{1}{3}\sin^2\theta_W D_{b2i}^{*}D_{b2l})\cos\beta \nonumber\\
-\frac{gm^2_b}{\sqrt 2 m_W \cos\beta}(D_{b1i}^{*}D_{b1l}+
D_{b2i}^{*}D_{b2l})-\frac{g m_b A_b}{\sqrt 2 m_W \cos\beta}
D_{b2i}^{*}D_{b1l}
\eeqn
The matrix elements $D_q$ are defined as
\beq
D^{+}_q M^2_{\tilde{q}} D_q =diag (m^2_{\tilde{q1}},
m^2_{\tilde{q2}})
\eeq
We need also the $\bar{t}\chi^{+} \tilde{b}$ interaction which is given by
\beqn
{\cal{L}}_{\bar{t}\chi^{+}\tilde{b}}=
 -g\bar{\chi^{+}_k}[(U^{*}_{k1} D^{*}_{b_{1i}} -\kappa_b U^{*}_{k2} D^{*}_{b_{2i}})P_L\nonumber\\
-\kappa_t V_{k2} D^{*}_{b_{1i}}P_R]t \tilde{b}^{*}_i +H.c
\eeqn
where $\kappa_{t,b}$ are given by
\beqn
\kappa_t=\frac{m_t}{\sqrt 2 m_W \sin\beta}\nonumber\\
~\kappa_b=\frac{m_b}{\sqrt 2 m_W \cos\beta}
\eeqn
For external momenta $s$, $q$ and $q-s$ the amplitude correction from loop 1i(a)
is given by
\beqn
\delta M = -g^2 H_{il} \bar{u}(q-s)[C_{L_{jl}}P_L+C_{R_{jl}}P_R]\nonumber\\
\int \frac{d^4\ell}{(2\pi)^4}[(\not\!s+\not\!\ell)+m_t][C^{*}_{L_{ki}}P_R+C^{*}_{R_{ki}}P_L]v(s)\nonumber\\
\times \frac{1}{((s+\ell)^2-m^2_t+i\epsilon)
(\ell^2-m^2_{\tilde{b_l}}+i\epsilon)((\ell+q)^2-m^2_{\tilde{b_i}}+i\epsilon)}
\eeqn
where $C_{L_{jl}}$ and $C_{R_{jl}}$ are given by
\beqn
C_{L_{jl}}=U^{*}_{j1} D^{*}_{b_{1l}} -\kappa_b U^{*}_{j2} D^{*}_{b_{2l}}\nonumber\\
C_{R_{jl}}=-\kappa_t V_{j2}D^{*}_{b_{1l}}
\eeqn
The part in the numerator
\beqn
[C_{L_{jl}}P_L+C_{R_{jl}}P_R][(\not\!s+\not\!\ell)+m_t]\nonumber\\
(C^{*}_{L_{ki}}P_R+C^{*}_{R_{ki}}P_L)
\eeqn
%\beqn
%[C_{L_{jl}}P_L+C_{R_{jl}}P_R][(\not\!s-\not\!\ell)+m_t]\nonumber\\
%[C^{*}_{L_{ki}}P_R+C^{*}_{R_{ki}}P_L]
%\eeqn
could be written as
\beqn
[C_{L_{jl}}C^{*}_{L_{ki}}P_L+C_{R_{jl}}C^{*}_{R_{ki}}P_R](\not\!s+\not\!\ell)\nonumber\\
+m_t[C_{R_{jl}}C^{*}_{L_{ki}}P_R+C_{L_{jl}}C^{*}_{R_{ki}}P_L]
\label{very}
\eeqn
by using the facts that $\gamma^{\mu} P_L=P_R\gamma^{\mu}$, $P_L P_R=0$, 
$P^2_L=P_L$ and $P^2_R=P_R$.
The first term in Eq. (\ref{very}) does not contribute to $\delta \phi_{jk}$ or $\Delta \psi_{jk}$ since it does not have the same Lorentz structure.
The second term of Eq. (\ref{very}) contributes the part of $m_tC_{R_{jl}}C^{*}_{L_{ki}}$
to $\delta \phi_{jk}$ and $m_tC_{L_{jl}}C^{*}_{R_{ki}}$
to $\Delta \psi_{jk}$.
Thus the loop corrections $\delta \phi_{jk}$ and $\Delta \psi_{jk}$ read
\beqn
i\delta \phi_{jk}=-g^2H_{il}m_t C_{R_{jl}}C^{*}_{L_{ki}} J\nonumber\\
i\Delta \psi_{jk}=-g^2H_{il}m_t C_{L_{jl}}C^{*}_{R_{ki}} J\nonumber\\
\eeqn
where
\beqn
J=\int \frac{d^4\ell}{(2\pi)^4}\frac{1}{((s+\ell)^2-m^2_t+i\epsilon)
(\ell^2-m^2_{\tilde{b_l}}+i\epsilon)((\ell+q)^2-m^2_{\tilde{b_i}}+i\epsilon)}
\eeqn
Now for zero external momentum approximation we set $s=q=0$, and the integral would read
\beq
\int \frac{d^4\ell}{(2\pi)^4}\frac{1}{(\ell^2-m^2_t+i\epsilon)
(\ell^2-m^2_{\tilde{b_l}}+i\epsilon)(\ell^2-m^2_{\tilde{b_i}}+i\epsilon)}
\eeq
A detailed calculation of this integral is given in the appendix.

Using the above one finds for $\delta \phi_{jk}$ the contribution: 
\beqn
\delta\phi^{(1)}_{jk}=\kappa_t \frac{g^2 m_t}{16 \pi^2} \sum_{i=1}^2\sum_{l=1}^2 H_{il} V_{j2} D^{*}_{b_{1l}} (U_{k1} D_{b_{1i}} -\kappa_b U_{k2} D_{b_{2i}}) f(m^2_t, m^2_{\tilde{b}_l}, m^2_{\tilde{b}_i})
\eeqn
where
\beq
f(x,y,z)=\frac{1}{(x-y)(x-z)(z-y)} 
\times (zx ln\frac{z}{x}+xy ln\frac{x}{y}
+yz ln\frac{y}{z}),
\eeq
and
\beq
f(x,y,y)= \frac{1}{(y-x)^2}\times (xln\frac{y}{x}+x-y)
\eeq
Similarly one finds for the correction $\Delta\psi_{jk}$ from the same loop
the following contribution
\beqn
\Delta\psi^{(1)}_{jk}=\kappa_t \frac{g^2 m_t}{16 \pi^2} \sum_{i=1}^2\sum_{l=1}^2 H_{il} V^{*}_{k2} D_{b_{1i}} (U^{*}_{j1} D^{*}_{b_{1l}} -\kappa_b U^{*}_{j2} D^{*}_{b_{2l}}) f(m^2_t, m^2_{\tilde{b}_l}, m^2_{\tilde{b}_i})
\eeqn

Next for the loop Fig. 1ii(a) we find
\beqn
\delta\phi^{(2)}_{jk}=0\nonumber\\
~\Delta\psi^{(2)}_{jk}=0
\eeqn

For the loop of Fig. 1i(b) we find
\beqn
\delta\phi^{(3)}_{jk}=\kappa_b \frac{g^2 m_b}{16 \pi^2} \sum_{i=1}^2\sum_{l=1}^2
 F_{li} U_{k2} D^{*}_{t_{1i}} (V_{j1}D_{t_{1l}}-\kappa_tV_{j2}D_{t_{2l}}) f(m^2_b, m^2_{\tilde{t}_i}, m^2_{\tilde{t}_l})\nonumber\\
\Delta\psi^{(3)}_{jk}= \kappa_b \frac{g^2 m_b}{16 \pi^2} \sum_{i=1}^2\sum_{l=1}^2
 F_{li} U^{*}_{j2} D_{t_{1l}} (V^{*}_{k1}D^{*}_{t_{1i}}-\kappa_t V^{*}_{k2}D^{*}_{t_{2i}}) f(m^2_b, m^2_{\tilde{t}_i}, m^2_{\tilde{t}_l})
\eeqn
where $F_{li}$ is given by
\beqn
F_{li}=-\frac{gM_Z}{\sqrt 2 \cos\theta_W}
((\frac{1}{2}-\frac{2}{3}\sin^2\theta_W)D_{t1l}^{*}D_{t1i}
+\frac{2}{3}\sin^2\theta_W D_{t2l}^{*}D_{t2i})\cos\beta \nonumber\\
+\frac{g m_t \mu}{\sqrt 2 m_W \sin\beta}
D_{t1l}^{*}D_{t2i}
\eeqn

For the loop of Fig. 1ii(b) we find
\beqn
\delta\phi^{(4)}_{jk}=0\nonumber\\
\Delta\psi^{(4)}_{jk}= -\kappa_b \frac{g^2 m^2_b}{16 \pi^2}h_b \sum_{i=1}^2
  U^{*}_{j2} D_{t_{1i}} (V^{*}_{k1}D^{*}_{t_{1i}}-\kappa_t V^{*}_{k2}D^{*}_{t_{2i}}) f(m^2_b, m^2_b, m^2_{\tilde{t}_i})
\eeqn

For loop of Fig. 1ii(c) we find
\beqn
\delta\phi^{(5)}_{jk}=2g\sum_{i=1}^4\sum_{l=1}^4 Q^{'}_{il}
\epsilon^{'}_{ik}\sin\beta 
\epsilon^{*}_{lj}\cos\beta\nonumber\\
\frac{m_{\chi^0_i}m_{\chi^0_l}}{16 \pi^2}
f(m^2_{\chi^0_i},m^2_{\chi^0_l} , m^2_{H^+})\nonumber\\
\Delta\psi^{(5)}_{jk}=0
\eeqn
where $\epsilon^{'}$ and $\epsilon$ are given by
\beqn
\epsilon_{ji}=-gX_{4j}V_{i1}^{*}-\frac{g}{\sqrt 2}X_{2j}V_{i2}^{*}-\frac{g}{\sqrt 2} \tan\theta_W X_{1j}V_{i2}^{*}\nonumber\\
~\epsilon^{'}_{ji}=-gX_{3j}^{*}U_{i1}+\frac{g}{\sqrt 2}X_{2j}^{*}U_{i2} +\frac{g}{\sqrt 2} \tan\theta_W X_{1j}^{*}U_{i2}
\eeqn

The parameters $Q^{'}_{ij}$ are defined as:
\beq
Q^{'}_{ij}=\frac{1}{\sqrt 2}[X^{*}_{3i}(X^{*}_{2j}-\tan \theta_W
X^{*}_{1j})]
\eeq
The matrix elements $X$ are defined as
\beq
X^T M_{\chi^0} X=diag(m_{\chi^0_1}, m_{\chi^0_2}, m_{\chi^0_3}, m_{\chi^0_4})
\eeq
For loop of Fig. 1i(c) we find
\beqn
\delta\phi^{(6)}_{jk}= \frac{g m_W \cos\beta}{2 \sqrt 2}
[1+2 \sin^2\beta -\cos2\beta \tan^2\theta_W]\sum_{i=1}^4\nonumber\\
\epsilon^{'}_{ik}\sin\beta
\epsilon^{*}_{ij}\cos\beta\nonumber\\
\frac{m_{\chi^0_i}}{16 \pi^2}
f(m^2_{\chi^0_i},m^2_{H^+}, m^2_{H^+})\nonumber\\
\Delta\psi^{(6)}_{jk}= \frac{g m_W \cos\beta}{2 \sqrt 2}
[1+2 \sin^2\beta -\cos2\beta \tan^2\theta_W]\sum_{i=1}^4\nonumber\\
\epsilon_{ik}\cos\beta
\epsilon^{'*}_{ij}\sin\beta \nonumber\\
\frac{m_{\chi^0_i}}{16 \pi^2}
f(m^2_{\chi^0_i},m^2_{H^+}, m^2_{H^+})
\eeqn

For loop of Fig. 1i(d) we find
\beqn
\delta\phi^{(7)}_{jk}= g^3 \frac{m_Z \cos\beta}{8\sqrt 2 \cos\theta_W}\sum_{l=1}^3\sum_{m=1}^3\sum_{i=1}^2
((Y_{m1}-iY_{m3}\sin\beta)(3 Y_{l1}+iY_{l3}\sin\beta)\nonumber\\-2
(Y_{m2}-iY_{m3}\cos\beta)(Y_{l2}+iY_{l3}\cos\beta)
-4Y_{m2}(Y_{l1}-iY_{l3}\sin\beta)\tan\beta)\nonumber\\
(Q_{ki}(Y_{l1}+iY_{l3}\sin\beta)+S_{ki}(Y_{l2}+iY_{l3}\cos\beta))\nonumber\\
(Q_{ij}(Y_{m1}+iY_{m3}\sin\beta)+S_{ij}(Y_{m2}+iY_{m3}\cos\beta))\nonumber\\
\frac{m_{\chi^+_i}}{16 \pi^2} f(m^2_{\chi^+_i},m^2_{H^0_m}, m^2_{H^0_l})\nonumber\\
\Delta\psi^{(7)}_{jk}= g^3 \frac{m_Z \cos\beta}{8\sqrt 2 \cos\theta_W}\sum_{l=1}^3\sum_{m=1}^3\sum_{i=1}^2
((Y_{m1}-iY_{m3}\sin\beta)(3 Y_{l1}+iY_{l3}\sin\beta)\nonumber\\-2
(Y_{m2}-iY_{m3}\cos\beta)(Y_{l2}+iY_{l3}\cos\beta)
-4Y_{m2}(Y_{l1}-iY_{l3}\sin\beta)\tan\beta)\nonumber\\
(Q^{*}_{ik}(Y_{l1}-iY_{l3}\sin\beta)+S^{*}_{ik}(Y_{l2}-iY_{l3}\cos\beta))\nonumber\\
(Q^{*}_{ji}(Y_{m1}-iY_{m3}\sin\beta)+S^{*}_{ji}(Y_{m2}-iY_{m3}\cos\beta))\nonumber\\
\frac{m_{\chi^+_i}}{16 \pi^2} f(m^2_{\chi^+_i},m^2_{H^0_m}, m^2_{H^0_l})
\eeqn
where $Q_{ji}=-\frac{1}{\sqrt 2 g}\phi_{ij}$ and 
$S_{ji}=\frac{1}{\sqrt 2 g}\psi_{ij}$, and the matrix elements $Y$ are defined as $Y M^2_{Higgs}Y^T=diag(m^2_{H^0_1},m^2_{H^0_2},m^2_{H^0_3})$.

For loop of Fig. 1ii(d) we find
\beqn
\delta\phi^{(8)}_{jk}= -g^2\sum_{m=1}^3\sum_{i=1}^2\sum_{l=1}^2
\phi_{li}\nonumber\\
(Q_{li}(Y_{m1}+iY_{m3}\sin\beta)+S_{lj}(Y_{m2}+iY_{m3}\cos\beta))
(Q_{ki}(Y_{m1}+iY_{m3}\sin\beta)\nonumber\\
+S_{ki}(Y_{m2}+iY_{m3}\cos\beta))
\frac{m_{\chi^+_i}m_{\chi^+_l}}{16 \pi^2}
f(m^2_{\chi^+_i},m^2_{H^0_m},m^2_{\chi^+_l})\nonumber\\
\Delta\psi^{(8)}_{jk}=0
\eeqn

For loop of Fig. 1ii(e) we find
\beqn
\delta\phi^{(9)}_{jk}=0\nonumber\\
~\Delta\psi^{(9)}_{jk}= \frac{4 g^2}{\cos^2\theta_W}
\sum_{l=1}^2\sum_{i=1}^2 \phi_{li} R'_{jl} L'_{ik}
\frac{m_{\chi^+_i}m_{\chi^+_l}}{16 \pi^2}
f(m^2_{\chi^+_i},m^2_{Z_0},m^2_{\chi^+_l})
\eeqn
The parameters $L'$ and $R'$ are defined by
\beqn
L_{ij}^{'}=-V_{i1} V^{*}_{j1} -\frac{1}{2}V_{i2} V^{*}_{j2}
+\delta_{ij} \sin^2\theta_W\nonumber\\
~ R_{ij}^{'}=-U^{*}_{i1} U_{j1} -\frac{1}{2}U^{*}_{i2} U_{j2}
+\delta_{ij} \sin^2\theta_W
\eeqn

For loop of Fig. 1i(e) we find
\beqn
\delta\phi^{(10)}_{jk}= -\frac{\sqrt 2 g^3 m_Z \cos\beta}{\cos^3\theta_W}\sum_{i=1}^2  L'_{ji} R'_{ik}
\frac{m_{\chi^+_i}}{16 \pi^2}
f(m^2_{\chi^+_i},m^2_{Z_0},m^2_{Z_0})\nonumber\\
~\Delta\psi^{(10)}_{jk}= -\frac{\sqrt 2 g^3 m_Z \cos\beta}{\cos^3\theta_W}\sum_{i=1}^2  R'_{ji} L'_{ik}
\frac{m_{\chi^+_i}}{16 \pi^2}
f(m^2_{\chi^+_i},m^2_{Z_0},m^2_{Z_0})
\eeqn

For loop of Fig. 1ii(f) we find
\beqn
\delta\phi^{(11)}_{jk}=0\nonumber\\
~\Delta\psi^{(11)}_{jk}=-4\sqrt 2 g^3 \sum_{i=1}^4 \sum_{l=1}^4
Q"_{il}R^{*}_{lj}L_{ik} 
\frac{m_{\chi^0_i}m_{\chi^0_l}}{16 \pi^2}
f(m^2_{\chi^0_i},m^2_{W^+},m^2_{\chi^0_l})
\eeqn
where $L$, $R$ and $Q"$ are defined as
\beqn
L_{ij}=-\frac{1}{\sqrt 2}X^{*}_{4i} V^{*}_{j2} +X^{*}_{2i} V^{*}_{j1}\nonumber\\
~R_{ij}=\frac{1}{\sqrt 2}X_{3i} U_{j2} +X_{2i} U_{j1}\nonumber\\
~gQ^{"}=\frac{1}{2}(X^{*}_{3i}(g X^{*}_{2j}-g'X^{*}_{1j})+(i\leftrightarrow j))
\eeqn

For loop of Fig. 1i(f) we find
\beqn
\delta\phi^{(12)}_{jk}=-\frac{4 g^3 m_W \cos\beta}{\sqrt 2}
\sum_{i=1}^4 L^{*}_{ij} R_{ik} 
\frac{m_{\chi^0_i}}{16 \pi^2}
f(m^2_{\chi^0_i},m^2_{W^+},m^2_{W^+})\nonumber\\
~\Delta\psi^{(12)}_{jk}=-\frac{4 g^3 m_W \cos\beta}{\sqrt 2}
\sum_{i=1}^4 R^{*}_{ij} L_{ik} 
\frac{m_{\chi^0_i}}{16 \pi^2}
f(m^2_{\chi^0_i},m^2_{W^+},m^2_{W^+})
\eeqn

For loop of Fig. 1ii(g) we find
\beqn
\delta\phi^{(13)}_{jk}=0\nonumber\\
~\Delta\psi^{(13)}_{jk}=-g^2 h_{\tau}
\kappa_{\tau} U^{*}_{j2} V^{*}_{k1} \frac{m^2_{\tau}}{16 \pi^2}
f(m^2_{\tau},m^2_{\tau},m^2_{\nu_{\tau}})
\eeqn
where 
\beq
\kappa_{\tau}=\frac{m_{\tau}}{\sqrt 2 m_W \cos\beta}
\eeq

The loop corrections for $\delta \phi_{jk}$ and 
$\Delta \psi_{jk}$ are given by
\beqn
\delta\phi_{jk}=\sum_{n=1}^{13} \delta\phi^{(n)}_{jk}\nonumber\\
~\Delta\psi_{jk}=\sum_{n=1}^{13} \Delta\psi^{(n)}_{jk}
\eeqn

\subsection{Loop analysis of $\Delta \phi_{jk}$ and  $\delta \psi_{jk}$}

We do the same analysis of Fig. 2 as for Fig. 1. We write down here the final results for both corrections from the thirteen loops together. The corrections are written in the same order of the loops in Fig. 2.
\beqn
\Delta\phi_{jk}=\kappa_t \frac{g^2 m_t}{16 \pi^2} \sum_{i=1}^2\sum_{l=1}^2 G_{il} V^{*}_{k2} D_{b_{1i}} (U^{*}_{j1} D^{*}_{b_{1l}} -\kappa_b U^{*}_{j2} D^{*}_{b_{2l}}) f(m^2_t, m^2_{\tilde{b}_l}, m^2_{\tilde{b}_i})\nonumber\\
-\kappa_t h_t\frac{g^2 m^2_t}{16 \pi^2} \sum_{i=1}^2  V^{*}_{k2} D_{b_{1i}} (U^{*}_{j1} D^{*}_{b_{1i}} -\kappa_b U^{*}_{j2} D^{*}_{b_{2i}}) f(m^2_t, m^2_t, m^2_{\tilde{b}_i})\nonumber\\
+\kappa_b \frac{g^2 m_b}{16 \pi^2} \sum_{i=1}^2\sum_{l=1}^2
 E_{li} U^{*}_{j2} D_{t_{1l}} (V^{*}_{k1}D^{*}_{t_{1i}}-\kappa_t V^{*}_{k2}D^{*}_{t_{2i}}) f(m^2_b, m^2_{\tilde{t}_i}, m^2_{\tilde{t}_l})\nonumber\\
+0\nonumber\\
+0\nonumber\\
+\frac{g m_W \sin\beta}{2 \sqrt 2}
[1+2 \cos^2\beta +\cos2\beta \tan^2\theta_W]\sum_{i=1}^4\nonumber\\
\epsilon_{ik}\cos\beta
\epsilon^{'*}_{ij}\sin\beta\nonumber\\
\frac{m_{\chi^0_i}}{16 \pi^2}
f(m^2_{\chi^0_i},m^2_{H^+}, m^2_{H^+})\nonumber\\
+g^3 \frac{m_Z \cos\beta}{8\sqrt 2 \cos\theta_W}\sum_{l=1}^3\sum_{m=1}^3\sum_{i=1}^2
(\tan\beta(Y_{l2}-iY_{l3}\cos\beta)(3 Y_{m2}+iY_{m3}\cos\beta)\nonumber\\-4 Y_{l1}
(Y_{m2}-iY_{m3}\cos\beta)-2\tan\beta(Y_{m1}-iY_{m3}\sin\beta)(Y_{l1}+iY_{l3}\sin\beta))\nonumber\\
(Q^{*}_{ik}(Y_{l1}-iY_{l3}\sin\beta)+S^{*}_{ik}(Y_{l2}-iY_{l3}\cos\beta))\nonumber\\
(Q^{*}_{ji}(Y_{m1}-iY_{m3}\sin\beta)+S^{*}_{ji}(Y_{m2}-iY_{m3}\cos\beta))\nonumber\\
\frac{m_{\chi^+_i}}{16 \pi^2} f(m^2_{\chi^+_i},m^2_{H^0_m}, m^2_{H^0_l})\nonumber\\
+0\nonumber\\
+\frac{4 g^2}{\cos^2\theta_W}
\sum_{l=1}^2\sum_{i=1}^2 \psi_{li} R'_{jl} L'_{ik}
\frac{m_{\chi^+_i}m_{\chi^+_l}}{16 \pi^2}
f(m^2_{\chi^+_i},m^2_{Z_0},m^2_{\chi^+_l})\nonumber\\
-\frac{\sqrt 2 g^3 m_Z \sin\beta}{\cos^3\theta_W}\sum_{i=1}^2  R'_{ji} L'_{ik}
\frac{m_{\chi^+_i}}{16 \pi^2}
f(m^2_{\chi^+_i},m^2_{Z_0},m^2_{Z_0})\nonumber\\
-4\sqrt 2 g^3 \sum_{i=1}^4 \sum_{l=1}^4
S"_{il}R^{*}_{lj}L_{ik} 
\frac{m_{\chi^0_i}m_{\chi^0_l}}{16 \pi^2}
f(m^2_{\chi^0_i},m^2_{W^+},m^2_{\chi^0_l})\nonumber\\
-\frac{4 g^3 m_W \sin\beta}{\sqrt 2}
\sum_{i=1}^4 R^{*}_{ij} L_{ik} 
\frac{m_{\chi^0_i}}{16 \pi^2}
f(m^2_{\chi^0_i},m^2_{W^+},m^2_{W^+})\nonumber\\
+0
\eeqn
where $G$ and $E$ are given by
\beqn
G_{ij}=\frac{gM_Z}{\sqrt 2 \cos\theta_W}
((-\frac{1}{2}+\frac{1}{3}\sin^2\theta_W)D_{b1i}^{*}D_{b1j}
-\frac{1}{3}\sin^2\theta_W D_{b2i}^{*}D_{b2j})\sin\beta \nonumber\\
+\frac{g m_b \mu}{\sqrt 2 m_W \cos\beta}
D_{b1i}^{*}D_{b2j}\nonumber\\
E_{ij}=\frac{gM_Z}{\sqrt 2 \cos\theta_W}
((\frac{1}{2}-\frac{2}{3}\sin^2\theta_W)D_{t1i}^{*}D_{t1j}
+\frac{2}{3}\sin^2\theta_W D_{t2i}^{*}D_{t2j})\sin\beta \nonumber\\
-\frac{gm^2_t}{\sqrt 2 m_W \sin\beta}(D_{t1i}^{*}D_{t1j}+
D_{t2i}^{*}D_{t2j})-\frac{g m_t A_t}{\sqrt 2 m_W \sin\beta}
D_{t2i}^{*}D_{t2j}
\eeqn
and $S"$ is given by
\beqn
S^{"}_{li}=-\frac{1}{\sin\beta}(\frac{M_l}{2m_W}\delta_{li}
-Q^{"}_{li}\cos\beta -R^{"}_{li})\nonumber\\
R^{"}_{li}=\frac{1}{2m_W}
(\tilde{m_1}^{*} X_{1l}^{*} X_{1i}^{*}
+\tilde{m_2}^{*} X_{2l}^{*} X_{2i}^{*}
-\mu^{*}(X_{3l}^{*} X_{4i}^{*}+X_{4l}^{*} X_{3i}^{*}))
\eeqn

The corrections $\delta \psi_{jk}$ are given by
\beqn
\delta\psi_{jk}=\kappa_t \frac{g^2 m_t}{16 \pi^2} \sum_{i=1}^2\sum_{l=1}^2 G_{il} V_{j2} D^{*}_{b_{1l}} (U_{k1} D_{b_{1i}} -\kappa_b U_{k2} D_{b_{2i}}) f(m^2_t, m^2_{\tilde{b}_l}, m^2_{\tilde{b}_i})\nonumber\\
+0\nonumber\\
+\kappa_b \frac{g^2 m_b}{16 \pi^2} \sum_{i=1}^2\sum_{l=1}^2
 E_{li} U_{k2} D^{*}_{t_{1i}} (V_{j1}D_{t_{1l}}-\kappa_tV_{j2}D_{t_{2l}}) f(m^2_b, m^2_{\tilde{t}_i}, m^2_{\tilde{t}_l})\nonumber\\
+0\nonumber\\
-2g\sum_{i=1}^4\sum_{l=1}^4 S^{'}_{il}
\epsilon^{'}_{ik}\sin\beta
\epsilon^{*}_{lj}\cos\beta \nonumber\\
\frac{m_{\chi^0_i}m_{\chi^0_l}}{16 \pi^2}
f(m^2_{\chi^0_i},m^2_{\chi^0_l} , m^2_{H^+})\nonumber\\
+\frac{g m_W \sin\beta}{2 \sqrt 2}
[1+2 \cos^2\beta +\cos2\beta \tan^2\theta_W]\sum_{i=1}^4\nonumber\\
\epsilon^{'}_{ik}\sin\beta
\epsilon^{*}_{ij}\cos\beta \nonumber\\
\frac{m_{\chi^0_i}}{16 \pi^2}
f(m^2_{\chi^0_i},m^2_{H^+}, m^2_{H^+})\nonumber\\
+g^3 \frac{m_Z \cos\beta}{8\sqrt 2 \cos\theta_W}\sum_{l=1}^3\sum_{m=1}^3\sum_{i=1}^2
(\tan\beta(Y_{l2}-iY_{l3}\cos\beta)(3 Y_{m2}+iY_{m3}\cos\beta)\nonumber\\-4 Y_{l1}
(Y_{m2}-iY_{m3}\cos\beta)-2\tan\beta(Y_{m1}-iY_{m3}\sin\beta)(Y_{l1}+iY_{l3}\sin\beta))\nonumber\\
(Q_{ki}(Y_{l1}+iY_{l3}\sin\beta)+S_{ki}(Y_{l2}+iY_{l3}\cos\beta))\nonumber\\
(Q_{ij}(Y_{m1}+iY_{m3}\sin\beta)+S_{ij}(Y_{m2}+iY_{m3}\cos\beta))\nonumber\\
\frac{m_{\chi^+_i}}{16 \pi^2} f(m^2_{\chi^+_i},m^2_{H^0_m}, m^2_{H^0_l})\nonumber\\
-g^2\sum_{m=1}^3\sum_{i=1}^2\sum_{l=1}^2
\psi_{li}\nonumber\\
(Q_{lj}(Y_{m1}+iY_{m3}\sin\beta)+S_{lj}(Y_{m2}+iY_{m3}\cos\beta))
(Q_{ki}(Y_{m1}+iY_{m3}\sin\beta)\nonumber\\
+S_{ki}(Y_{m2}+iY_{m3}\cos\beta))
\frac{m_{\chi^+_i}m_{\chi^+_l}}{16 \pi^2}
f(m^2_{\chi^+_i},m^2_{H^0_m},m^2_{\chi^+_l})\nonumber\\
+0\nonumber\\
-\frac{\sqrt 2 g^3 m_Z \sin\beta}{\cos^3\theta_W}\sum_{i=1}^2  L'_{ji} R'_{ik}
\frac{m_{\chi^+_i}}{16 \pi^2}
f(m^2_{\chi^+_i},m^2_{Z_0},m^2_{Z_0})\nonumber\\
+0\nonumber\\
-\frac{4 g^3 m_W \sin\beta}{\sqrt 2}
\sum_{i=1}^4 L^{*}_{ij} R_{ik} 
\frac{m_{\chi^0_i}}{16 \pi^2}
f(m^2_{\chi^0_i},m^2_{W^+},m^2_{W^+})\nonumber\\
+0
\eeqn
where $S'$ is given by
\beq
S^{'}_{ij}=\frac{1}{\sqrt 2}[X^{*}_{4j}(X^{*}_{2i}-\tan \theta_W
X^{*}_{1i})]
\eeq

\section{Neutral Higgs decays including loop effects}

We summarize now the result of the analysis. 
Thus ${\cal{L}}_{eff}$ of $Eq.(5)$ may be written as follows
\begin{equation}
{\cal {L}}_{eff}=H^{0}_{l}\overline{\chi^+_j}(\alpha^{lS}_{jk}+\gamma_{5}\alpha^{lP}_{jk})\chi^+_k+ 
H.c
\end{equation}
where
\begin{equation}
\alpha^{lS}_{jk}=\frac{1}{2\sqrt 2}((Y_{l1}+iY_{l3}\sin\beta)(\phi_{jk}+\delta \phi_{jk}+\Delta \psi_{jk})+
(Y_{l2}+iY_{l3}\cos\beta)(\psi_{jk}+\delta \psi_{jk}+\Delta \phi_{jk}))
\label{scoupling}
\end{equation}
and where
\begin{equation}
\alpha^{lP}_{jk}=\frac{1}{2\sqrt 2}((Y_{l1}+iY_{l3}\sin\beta)(\phi_{jk}+\delta \phi_{jk}-\Delta \psi_{jk})+
(Y_{l2}+iY_{l3}\cos\beta)(\psi_{jk}+\delta \psi_{jk}-\Delta \phi_{jk}))
\label{pcoupling}
\end{equation}
Next we discuss the implications of the above result for the 
decay of the neutral Higgs.

\beqn
\Gamma_{ljk}(H^{0}_l\rightarrow\chi^+_j\chi^-_k)=\frac{1}{4\pi 
M^3_{H^0_l}}
\sqrt{[(m^2_{\chi^+_j}+m^2_{\chi^{+}_k}-M^2_{H^{0}_l})^2
-4m^2_{\chi^{+}_k}m^2_{\chi^+_j}]}\nonumber\\
([\frac{1}{2}((|\alpha^{lS}_{jk}|)^2+(|\alpha^{lP}_{jk}|)^2)
(M^2_{H^{0}_l}-m^2_{\chi^{+}_k}-m^2_{\chi^{+}_j})
-\frac{1}{2}((|\alpha^{lS}_{jk}|)^2-(|\alpha^{lP}_{jk}|)^2)
(2m_{\chi^{+}_k}m_{\chi^{+}_j})])
\label{branching}
\eeqn

There are many channels for $H^0_l$ decays. 
The important channels for the decay of the  neutral Higgs boson are
$\bar{b} b$, $\bar{t} t$, $\bar{s} s$, $\bar{c} c$, $\bar{\tau} \tau$, 
 $\chi^{+}_i \chi^{-}_j$ and $\chi^0_i \chi^0_j$. 
There is another set of channels that neutral Higgs can also decay into: 
these are modes
of decaying into the other fermions of the SM,  squarks, sleptons,
other Higgs bosons, W and Z boson pairs, one Higgs and a vector boson, 
$\gamma \gamma$ pairs and finally into the gluonic decay i.e, $H^0_l\rightarrow g g$.
We neglect the lightest SM fermions for the smallness 
of their couplings.
We choose the region in the parameter space where we can ignore
the other channels which either are not allowed kinematically or suppressed by 
their couplings. Thus in this work, squarks and sleptons are too heavy to be
relevant in neutral Higgs decay. The neutral Higgs decays into
nonsupersymmetric final states that involve gauge bosons and/or 
other Higgs bosons are ignored as well.
In the region of large $\tan\beta$, 
these decays typically contribute less than $1\%$ of the
total Higgs decay rate \cite{gunion}.
Thus we can neglect these final states.

We calculate the radiative corrected partial decay widths of the important channels
mentioned above. In the case of CP violating case under investigation we 
use for the radiatively corrected $\Gamma$ of neutral Higgs into quarks and leptons the analysis of \cite{we1}, for the radiatively corrected partial widths into
charginos we use the current analysis, and for the radiatively  corrected decay width  into neutralino
we use \cite{we6}.
We define 
\beq
\Delta\Gamma_{l}^{i,j}=\frac{\Gamma(H^0_l\rightarrow\chi_i^+\chi_j^-)-\Gamma^0(H^0_l\rightarrow\chi_i^+\chi_j^-)}{\Gamma^0(H^0_l\rightarrow\chi_i^+\chi_j^-)}
\label{gammad}
\eeq
where the first term in the numerator is the decay width including
the full loop corrections and the second term is the decay width
evaluated at the tree level.
Finally to quantify the size of the loop effects on 
the  branching  ratios of the neutral Higgs decay we define 
the following quantity
\beq
\Delta Br_{l}^{i,j}=\frac{Br(H^0_l\rightarrow\chi_i^+\chi_j^-)-Br^0(H^0_l\rightarrow\chi_i^+\chi_j^-)}{Br^0(H^0_l\rightarrow\chi_i^+\chi_j^-)}
\label{bran1}
\eeq
where the first term in the numerator is the branching ratio including
the full loop corrections and the second term is the branching ratio
evaluated at the tree level.
The analysis of this section is utilized in Sec.(4) where we 
give a numerical analysis of the size of the loop effects and
discuss the effect of the loop corrections on decay widths and branching 
ratios.

\section{NUMERICAL ANALYSIS}

In this section we discuss in a quantitative fashion the size of loop
effects on the partial decay width and the branching ratios of the neutral Higgs bosons into charginos.
The analysis of Sec. 2 is quite general and valid for the minimal 
supersymmetric standard model. For the sake of numerical analysis we will limit 
the parameter space by working within the framework of the SUGRA model 
\cite{sugra1}. Specifically we will work within the framework of the the extended
mSUGRA model including CP phases. 
We take as our parameter space at the grand unification scale to be 
the following: the universal scalar mass $m_0$, the universal gaugino mass $m_{1/2}$, 
the universal trilinear coupling 
$|A_0|$, the ratio of the Higgs vacuum expectation values $\tan\beta=<H_2>/<H_1>$
where $H_2$ gives mass to the up quarks and $H_1$ gives mass to the down 
quarks and the leptons.
In addition, we take for CP phases the following: the phase $\theta_{\mu}$ of the Higgs mixing parameter $\mu$, the phase $\alpha_{A_0}$ of the trilinear coupling $A_0$ and the phases $\xi_i(i=1,2,3)$ of the $SU(3)_C$, $SU(2)_L$ and $U(1)_Y$ gaugino masses.
In this analysis the electroweak symmetry is broken
by radiative effects which allows one to determine the
magnitude of $\mu$ by fixing $M_Z$. In the analysis we use
one loop renormalization group (RGEs) equations for the evolution
of the soft susy breaking parameters and for the parameter $\mu$, and two loop RGEs for the gauge and Yukawa couplings.
In the numerical analysis we compute the loop corrections and also analyze 
their dependence on the phases. The masses of particles involved in the analysis
are ordered as follows: 
for charginos $m_{\chi^+_1}<m_{\chi^+_2}$ and
for the neutral Higgs $(m_{H_1},m_{H_2},m_{H_3})\rightarrow (m_H,m_h,m_A)$ in the limit of no CP mixing where $m_H$ is the heavy CP even Higgs,
$m_h$ is the light CP even Higgs, and $m_A$ is the CP odd Higgs.

We investigate the question of how large loop corrections are relative to the tree
values. 
 We first discuss the magnitude of the loop corrections of the
partial decay width defined in Eq.(\ref{gammad}).
As we mentioned earlier the loop corrections  to the partial decay
width of the chargino channel have been investigated before
in the CP conserving case \cite{eberl,ren}.
The correction in these analyses is of  the order of $\sim 10\%$ of the tree level value. Our analysis supports
this conclusion. In Figs. (\ref{gamma1}) and (\ref{gamma3}) we give a plot of $\Delta \Gamma^{1,1}_l (l=1,3)$ as a function
of $\tan\beta$ for the specific set of inputs given in 
the captions of these figures. We notice that the 
partial decay width gets a change of $7\sim 15\%$ of
its tree level value. We also notice 
that the CP violating phase $\theta_{\mu}$ can affect the 
magnitude of this change. This effect has not been addressed in the previous
analyses as they are working in the CP conserving scenario. To compare between our analysis and the previous ones we have to notice that these analyses are using
 the general SUSY parameter space where they put by
hand all the parameters that control the analysis.
In \cite{eberl}, the authors choose the SUSY parameter set
SPS1a of the Snowmass Points and Slopes as a reference point.
They choose for the trilinear couplings the values of
$A_t=-487$ GeV, $A_b=-766$ GeV and $A_{\tau}=-250$ GeV. The
values of the other parameters are: 
$M=197.6$ GeV, $M'=98$ GeV, $\mu=353.1$ GeV, $\tan\beta=10$,
$m_{A^0}=393.6$ GeV, 
$M_{\tilde{Q}_{1,2}}=558.9$ GeV, $M_{\tilde{U}_{1,2}}=540.5$ GeV, $M_{\tilde{D}_{1,2}}=538.5$ GeV, 
$M_{\tilde{L}_{1,2}}=197.9$ GeV, $M_{\tilde{E}_{1,2}}=137.8$ GeV, $M_{\tilde{Q}_{3}}=512.2$ GeV, $M_{\tilde{U}_{3}}=432.8$ GeV, $M_{\tilde{D}_{3}}=536.5$ GeV,
$M_{\tilde{L}_{3}}=196.4$ GeV 
and $M_{\tilde{E}_{3}}=134.8$ GeV. In all the figures of \cite{eberl}, these
values are used, if not specified otherwise.
In our mSUGRA analysis the magnitude of all these parameters and
others are fixed by the five input parameters $m_0=100$ GeV, $m_{1/2}=250$ GeV,
$\tan\beta=10$, $A_0=-100$ GeV and a positive sign of $\mu$ in the CP conserving scenario \cite{allan}.
These parameters are different from those of our 
Figs. (\ref{gamma1}) and (\ref{gamma3}). 
By using these parameters and fixing some of them by hand when needed
to match their values in the analysis of \cite{eberl},
 we were able to have a fair agreement with their
Figs. (2-9).
 As an example of this check we show in Table.\ref{ya1} a comparison
of the two works. For the input of Fig. 2 of \cite{eberl} with CP violating phases
are set to zero we can see that partial decay widths in both works have the
same behavior as functions of masses and their magnitudes are fairly close to 
each other. However it seems that our loop corrected values of the partial widths are
different from those of Eberl et al. This could be understood since our loop analysis
of the effective lagrangian includes only the vertex corrections beside the 
corrections in the Higgs potential.

\begin{table}[h]
\begin{center}
\begin{tabular}{|r|r|r|r|r|}
\hline   
case & $\Gamma^{tree}_{eberl}$ & $\Gamma^{tree}_{our}$ & $\Gamma^{loop}_{eberl}$ & $\Gamma^{loop}_{our}$ \\ \hline       
2.a $m_{A_0}=700$ GeV  & $0.95$ GeV  & $0.94$ GeV  & $0.85$ GeV  & $0.80$ GeV \\ \hline
2.a $m_{A_0}=800$ GeV  & $1.18$ GeV  & $1.17$ GeV  & $1.0$ GeV  & $0.91$ GeV \\ \hline
2.b $m_{H_0}=800$ GeV  & $0.7$ GeV  & $0.69$ GeV  & $0.63$ GeV  & $0.58$ GeV \\ \hline
2.b $m_{H_0}=900$ GeV  & $0.8$ GeV  & $0.8$ GeV  & $0.73$ GeV  & $0.70$ GeV \\ \hline
\end{tabular}
\end{center}
\caption{A comparison between the current analysis and Eberl et al \cite{eberl} for benchmark cases. }
\label{ya1}
\end{table} 

In the work of Ref. \cite{ren} only 8 out of 26 diagrams of the present
analysis are calculated and they correspond to the 
vertex corrections from 
Figs. (1,2ii(a)), (1,2ii(b)), (1,2i(b)) and (1,2i(a)). 
By considering these diagrams only in the comparison,
our analysis is in fair agreement with their Figs (2-4) and Figs. 
(6,8) for their inputs.

Now we turn to address the question of how much loop corrections
can affect the branching ratios into charginos. The branching
ratio of a decay mode is defined to be the ratio between
the partial decay rate of this mode and the total decay rate.
In the parameter space under investigation this total decay
rate includes the rates of decays into charginos, heavy quarks, taus and neutralinos.
In Figs. (\ref{epsfig3a}) and (\ref{epsfig3c}) we give a plot of $\Delta Br^{1,1}_l (l=1,3)$ defined by Eq.(\ref{bran1}) as a function
of $\tan\beta$ for the specific set of inputs given in the captions of these figures. 
Fig. (\ref{epsfig3a}) is for the neutral Higgs $H_1$ boson and
Fig. (\ref{epsfig3c}) is for the neutral Higgs $H_3$ boson.
In all regions of the parameter space investigated in this work, the decay
of the lightest Higgs boson $H_2$ into charginos is forbidden
kinematically, since we have in these regions the fact that $2 m_{\chi^{-}_1}>m_{H_2}$.
The analysis of Figs. (\ref{epsfig3a}) and (\ref{epsfig3c})  shows that the loop correction varies
strongly with $\tan\beta$ with the correction changing sign for the case
of $H_3$ decay. Further, the analysis shows that the loop correction can be
as large as about $-40\%$ of the tree contribution for both $H_1$ and $H_3$ cases.
We also notice that the behavior of $\Delta Br^{1,1}_l (l=1,3)$ 
as a function of $\tan\beta$ changes considerably by
changing the phase of $\mu$. So for some values of this phase
we find that this parameter increases as $\tan\beta$ increases
and for other values of $\theta_{\mu}$ we see that it 
decreases as $\tan\beta$ increases.
As shown in the previous figures, the parameter $\tan\beta$ is playing a strong role.
This parameter is important at the tree level through the diagonalizing mass
matrices of the chargino and neutral Higgs and their spectrum. At the loop level it has extra
effect explicitly in $\alpha^{lP,S}_{jk}$ and implicitly through the radiatively
corrected matrix elements $Y_{lm}$
 and through the corrections
$\delta \phi_{jk}$, $\Delta \phi_{jk}$, $\delta \psi_{jk}$, $\Delta \psi_{jk}$. 
The values of the branching ratios themselves at tree and one loop levels are shown in Table.\ref{ya2}.

\begin{table}[h]
\begin{center}
\begin{tabular}{|r|r|r|r|r|}
\hline   
$\theta_{\mu}(rad)$ & $Br^0(H_1)$ & $Br^{loop}(H_1)$ & $Br^0(H_3)$ & $Br^{loop}(H_3)$ \\ \hline       
$0.5$  & $6\%$  & $4.7\%$  & $18.2\%$  & $13.8\%$ \\ \hline
$1.0$  & $8.4\%$  & $6.9\%$  & $21.3\%$  &  $18.1\%$ \\ \hline
$1.5$  & $9.2\%$  & $7.9\%$  & $23.4\%$  &  $22.2\%$ \\ \hline
\end{tabular}
\end{center}
\caption{Values of branching ratios at tree and one-loop levels of neutral Higgs into
the channel $\chi^+_1\chi^-_1$ at $\tan\beta=24$ for the input of Figs. (\ref{epsfig3a})
and (\ref{epsfig3c})}
\label{ya2}
\end{table}

We notice that their magnitudes are not negligible for the region
of the parameter space investigated. These non negligible branching ratios for the decay
of the neutral
Higgs  into charginos suggest that these decay modes
could be measurable at the soon-to-operate LHC. 
However, one should also consider the production rates for
 $H_1$ and $H_3$ bosons
to assess whether the change in branching ratios could be detectable at
colliders. This analysis goes beyond the scope of the current work.
We also notice
that the phase of the parameter $\mu$ affects the tree level branching ratios
as well. This comes mainly from the structure of the chargino matrix.
  The more important channels in the region of the parameter space
investigated are the decay into bottom and top quarks. They have the highest values of
branching ratios. The radiative corrections of these channels are
also more than those of the charginos and neutralinos.  These channels 
were studied  before  \cite{carena1,babu,we1} as mentioned above.
However a $20\%$ of branching ratio for the case of neutral Higgs
 as shown in the above table
is not very small and could justify carrying out the current analysis.

In Figs. (\ref{epsfig4a}) and (\ref{epsfig4c}) we give a plot of $\Delta Br^{1,1}_l (l=1,3)$ as a function of $|A_0|$ for the 
specific set of inputs given in the caption of these figures. The analysis of these figures shows
that the loop corrections are substantial and reaches the value of $-38\%$ of the 
tree contribution for the case of $H_1$ decay and the value
of $-43\%$ for the case of $H_3$ decay. 

Next we investigate the effects of CP violating phases on the 
loop corrections of the neutral Higgs decays into charginos.
In Figs. (\ref{epsfig5a}) and (\ref{epsfig5c}) we give a plot of 
$\Delta Br^{1,1}_l (l=1,3)$ as a function of $\theta_{\mu}$ for the 
specific set of inputs given in the caption of these figures. 
The analysis of the figures shows that the loop correction
has a sharp dependence on $\theta_{\mu}$. Further, the 
correction is changing sign as $\theta_{\mu}$ varies from 
$0$ to $\pi$ for two cases of  $H_3$ decay. Thus
$\theta_{\mu}$ affects not only the magnitude of $\Delta Br^{1,1}_l$ but also its sign depending on the value of
$\theta_{\mu}$. 

In Figs. (\ref{epsfig6a}) and (\ref{epsfig6c}) we give a plot of $\Delta Br^{1,1}_l (l=1,3)$ as a function of $\alpha_{A_0}$ for the specific set  
of inputs given in the caption of these figures. Here also we find 
a very substantial dependence of $\Delta Br^{1,1}_l$ on $\alpha_{A_0}$. This dependence is very large in the case of 
$H_3$ decay and it exceeds $-40\%$ of the tree contribution.

In Figs. (\ref{epsfig7a}) and (\ref{epsfig7c}) we give a plot of 
$\Delta Br^{1,1}_l (l=1,3)$ as a function of $\xi_2$ for the specific set  
of inputs given in the caption of these figures. Here we find
a small effect of this phase on the loop corrections. 
 
\section{CONCLUSION}

In this paper we have carried out an analysis of the supersymmetric loop corrections to $\chi^{+}_j\chi^{-}_kH^0_l$ couplings within MSSM. In supersymmetry after spontaneous
breaking of electroweak symmetry one is left with three 
neutral Higgs bosons which in the absence of CP phases consist
of two CP even Higgs bosons and one CP odd Higgs boson.
In the absence of loop corrections, the lightest Higgs 
boson mass satisfies the inequality $m_h<M_Z$ and by including
these corrections the lightest Higgs mass can be lifted
above $M_Z$.
 With
the inclusion of CP phases the Higgs boson mass eigenstates are
no longer CP even and CP odd states when loop corrections to the
Higgs boson mass matrix are included.
Further, inclusion of loop corrections to the couplings of charginos
and neutral Higgs is in general dependent on CP phases. Thus
the decays of neutral Higgs into charginos can be sensitive
to the loop corrections and to the CP violating phases.
The effect of the supersymmetric loop corrections is found to
to be in the range of $7\sim15\%$ for the partial decay 
width. For the branching ratios it is found to be
 be rather large, as much as $40\%$ in some regions of the
parameter space. The effect of CP phases on the modifications
of the partial decay width and the branching ratio is found to be substantial in some
regions of the MSSM parameter space.

\noindent
{\large\bf Acknowledgments}\\ 
I wish to acknowledge useful discussions with Professor Pran Nath. The support of the Physics Department at Alexandria
University is also acknowledged.

\section{APPENDIX}
The integral of import to this work is
\beq
J=\int \frac{d^4k}{(2\pi)^4}\frac{1}{(k^2-m^2_1+i\epsilon)
(k^2-m^2_2+i\epsilon)(k^2-m^2_3+i\epsilon)}
\eeq
It could be written in the form
\beq
\int \frac{d^4k}{(2\pi)^4} \frac{1}{D}
\eeq
where 
\beqn
\frac{1}{D}=\frac{1}{a}\frac{1}{b}\frac{1}{c}\nonumber\\
a=k^2-m^2_1+i\epsilon\nonumber\\
b=k^2-m^2_2+i\epsilon\nonumber\\
c=k^2-m^2_3+i\epsilon
\eeqn
Using Feynman parametrization, $\frac{1}{D}$ could be written as
\beq
\frac{1}{D}=2\int_0^1dx\int_0^{1-x}dz\frac{1}{[a+(b-a)x+(c-a)z]^3}
\eeq
The denominator in the above integral could be written in the form
$k^2+M^2+i\epsilon$ where $M^2=(m^2_1-m^2_2)x+(m^2_1-m^2_3)z-m^2_1$.
Thus the integral $J$ can take the form
\beq
J=\int \frac{d^4k}{(2\pi)^4} 2\int_0^1 dx \int_0^{1-x} dz \frac{1}{[k^2+M^2+i\epsilon]^3}
\eeq
Now integrating over $k$ and using the standard integral, for $n\geq 3$
\beq
\int \frac{d^4k}{(2\pi)^4}\frac{1}{(k^2+\Lambda+i\epsilon)^n}=i\pi^2\frac{\Gamma(n-2)}{\Gamma(n)}\frac{1}{\Lambda^{n-2}}
\eeq
one can find that the integral $J$ has the form
\beq
J=\frac{i}{(4\pi)^2}\int_0^1 dx \int_0^{1-x} dz \frac{1}{\alpha+\beta z}
\eeq
where $\alpha=(m^2_1-m^2_2)x-m^2_1$ and $\beta=m^2_1-m^2_3$.
Integrating over $z$ one can get for the integral $J$ the form of
\beq
J=\frac{i}{(4\pi)^2} \frac{1}{m^2_1-m^2_3}\int_0^{1} dx \ln(\delta_1 x -m^2_3)-\ln(\delta_2 x -m^2_1)
\eeq
where $\delta_1=m^2_3-m^2_2$ and $\delta_2=m^2_1-m^2_2$.
Finally we integrate over $x$ to get for $J$ the form of
\beq
J=\frac{i}{(4\pi)^2} f(m^2_1,m^2_2,m^2_3)
\eeq
where
\beqn
f(m^2_1,m^2_2,m^2_3)=\frac{1}{m^2_1-m^2_3} \frac{1}{m^2_3-m^2_2} \frac{1}{m^2_1-m^2_2}\nonumber\\
\times [m^2_2m^2_3\ln(\frac{m^2_2}{m^2_3})+
m^2_3m^2_1\ln(\frac{m^2_3}{m^2_1})+m^2_1m^2_2\ln(\frac{m^2_1}{m^2_2})]
\eeqn
This is the famous form factor that appears in the analysis of the radiative
corrections for the quark and lepton masses \cite{last1},
 the decay rates of neutral and charged Higgs into quarks and leptons \cite{carena1,babu,we1,we2}
and in the $b\rightarrow s \gamma$ process \cite{two}.
 In the latter process, the authors are using different
 form factor $H(\frac{m^2_1}{m^2_3},\frac{m^2_2}{m^2_3})$. This form factor
 could be easily converted to our
$f(m^2_1,m^2_2,m^2_3)$ through the simple relation
\beq
m^2_3 f(m^2_1,m^2_2,m^2_3)=H(\frac{m^2_1}{m^2_3},\frac{m^2_2}{m^2_3})
\eeq

For the case where two of the masses are equal, $m_2=m_3$, one can repeat the same 
analysis with $b=c=k^2-m^2_3+i\epsilon$ and $a=k^2-m^2_1+i\epsilon$. By doing so one can get
for the form factor $J$ 
\beq
J=\frac{i}{(4\pi)^2} \frac{1}{(m^2_3-m^2_1)^2} [m^2_1\ln(\frac{m^2_3}{m^2_1})+m^2_1-m^2_3]
\eeq

%%%%%%%%%%%%%%%%%%%%%%%%%%%%%%%%%%%%%%%%%%%%%%%%%%%%%%%%
\noindent
\begin{figure}[t]
\hspace*{-0.8in}
\centering
\includegraphics[width=18cm,height=18cm]{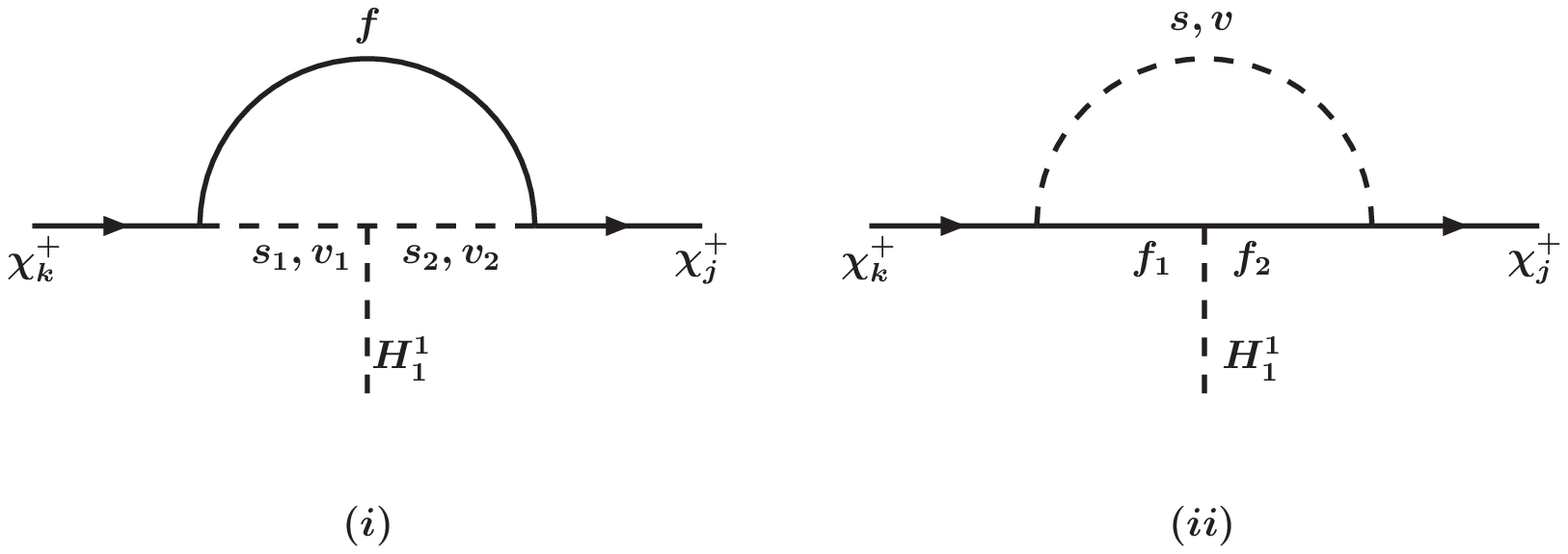}
\caption{Set of diagrams contributing to 
radiative corrections 
$\delta \phi_{jk}$ and $\Delta \psi_{jk}$.
 (i): (a) $s_1=\tilde{b}_{i}^{*}$, $s_2=\tilde{b}_{l}^{*}$, $f=t$; (b) $s_1=\tilde{t}_{i}$, $s_2=\tilde{t}_{l}$, $f=\bar{b}$; (c)
$s_1=H^+$, $s_2=H^+$, $f=\chi^0_i$; (d) $s_1=H^0_l$,
$s_2=H^0_m$, $f=\chi^{+}_i$; (e) $v_1=Z^0$, $v_2=Z^0$,
 $f=\chi^{+}_i$; (f) $v_1=W^+$, $v_2=W^+$, $f=\chi^{0}_i$.
(ii): (a) $f_1=t$, $f_2=t$, $s=\tilde{b}_{i}^{*}$;
(b) $f_1=\bar{b}$, $f_2=\bar{b}$, $s=\tilde{t}_{i}$;
(c)$f_1=\chi^0_i$, $f_2=\chi^0_l$, $s=H^+$;
(d) $f_1=\chi^+_i$, $f_2=\chi^+_l$, $s=H^0_m$;
(e) $f_1=\chi^+_i$, $f_2=\chi^+_l$, $v=Z^0$;
(f) $f_1=\chi^0_i$, $f_2=\chi^0_l$, $v=W$;
(g) $f_1=\tau^+$, $f_2=\tau^+$, $s=\tilde{\nu}_{\tau}$.}
\label{epsfig1}
\end{figure}

\begin{figure}[t]
\hspace*{-0.8in}
\centering
\includegraphics[width=18cm,height=18cm]{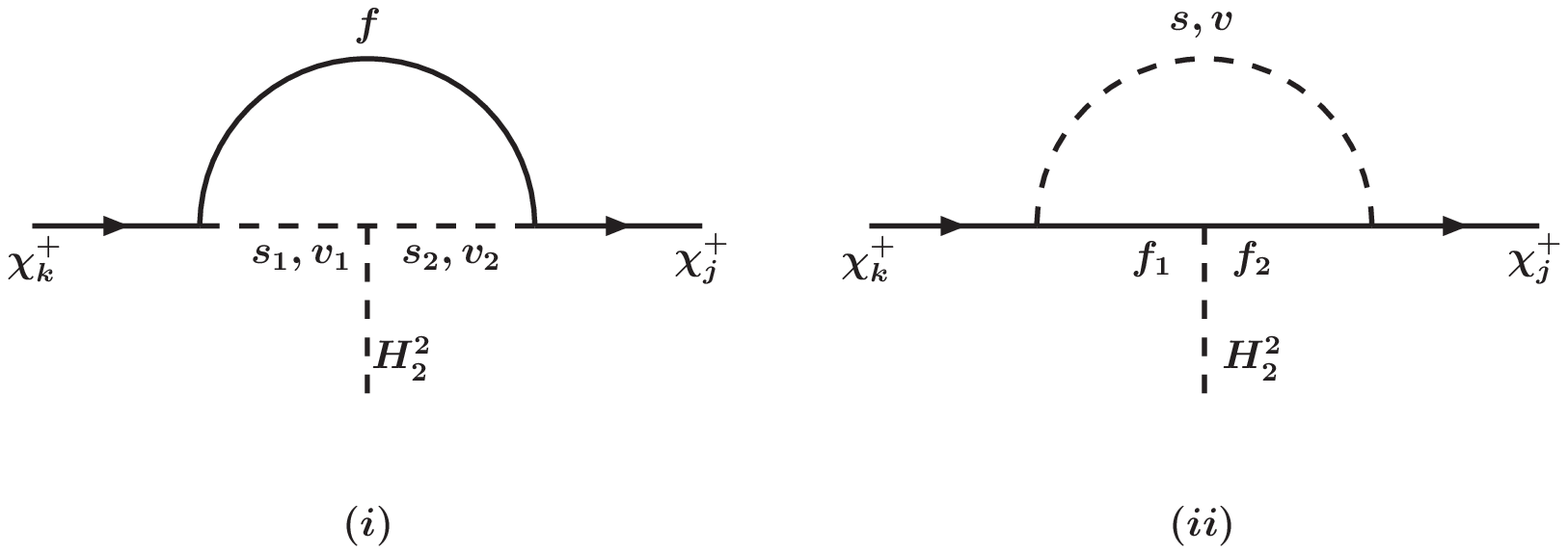}
\caption{Set of diagrams contributing to radiative corrections
$\Delta \phi_{jk}$ and $\delta \psi_{jk}$.
(i): (a) $s_1=\tilde{b}_{i}^{*}$, $s_2=\tilde{b}_{l}^{*}$, $f=t$; (b) $s_1=\tilde{t}_{i}$, $s_2=\tilde{t}_{l}$, $f=\bar{b}$; (c)
$s_1=H^+$, $s_2=H^+$, $f=\chi^0_i$; (d) $s_1=H^0_l$, $s_2=H^0_m$,
 $f=\chi^+_i$; (e) $v_1=Z^0$, $v_2=Z^0$, $f=\chi^+_i$; (f) $v_1=W^+$, $v_2=W^+$, $f=\chi^0_i$.
(ii): (a) $f_1=t$, $f_2=t$, $s=\tilde{b}_{i}^{*}$;
(b) $f_1=\bar{b}$, $f_2=\bar{b}$, $s=\tilde{t}_{i}$;
(c)$f_1=\chi^0_i$, $f_2=\chi^0_l$, $s=H^+$;
(d) $f_1=\chi^+_i$, $f_2=\chi^+_l$, $s=H^0_m$;
(e) $f_1=\chi^+_i$, $f_2=\chi^+_l$, $v=Z^0$;
(f) $f_1=\chi^0_i$, $f_2=\chi^0_l$, $v=W$;
(g) $f_1=\tau^+$, $f_2=\tau^+$, $s=\tilde{\nu}_{\tau}$.}
\label{epsfig2}
\end{figure}

\begin{figure}[t]
\hspace*{-0.8in}
\centering
\includegraphics[width=18cm,height=18cm]{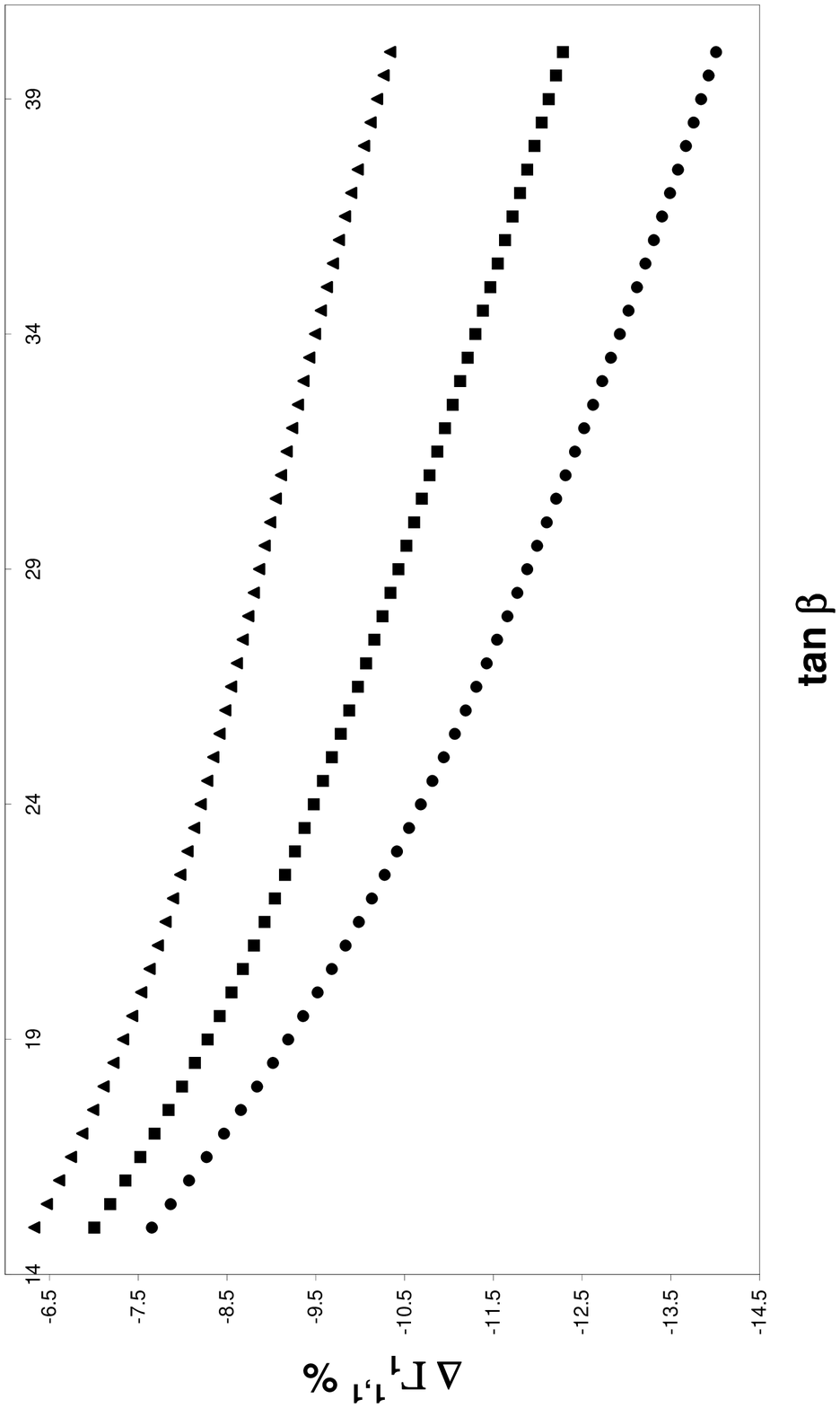}
\caption{$\tan\beta$ dependence of $\Delta\Gamma_1\to \chi_1^+\chi_1^-$.
The curves in ascending order correspond to $\theta_{\mu}=0.2$, $0.4$, $0.6$ (rad).
 The input is $m_0=350$ GeV,
$m_{1/2}=180$ GeV,
 $\xi_1=0.4$ (rad), $\xi_2=0.5$ (rad), $\xi_3=0.6$ (rad), $\alpha_{A_0}=0.8$ (rad) and $|A_0|=250$ GeV.}
\label{gamma1}
\end{figure}

\begin{figure}[t]
\hspace*{-0.8in}
\centering
\includegraphics[width=18cm,height=18cm]{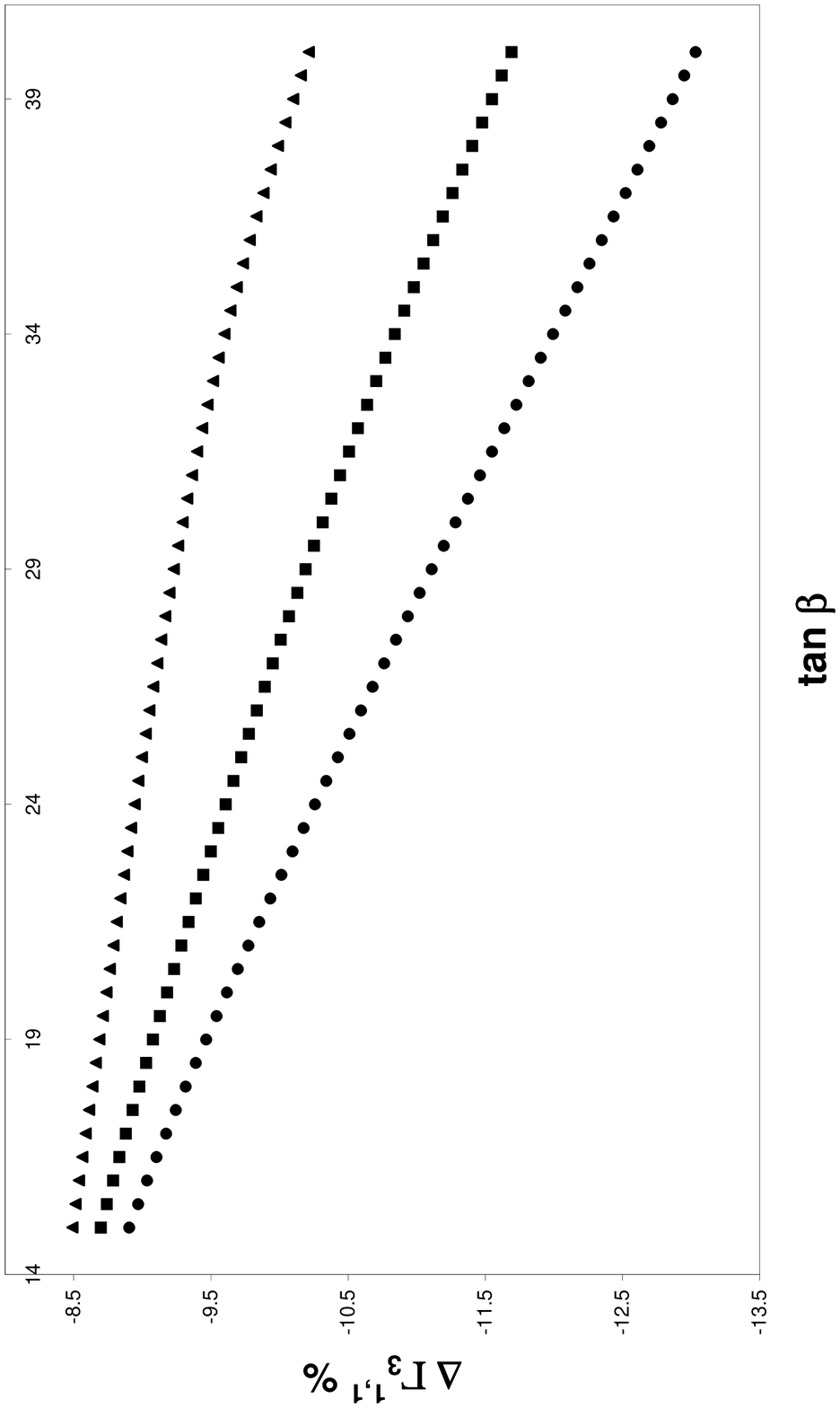}
\caption{$\tan\beta$ dependence of $\Delta\Gamma_3\to \chi_1^+\chi_1^-$.
The curves in ascending order correspond to $\theta_{\mu}=0.2$, $0.4$, $0.6$ (rad).
 The input is $m_0=350$ GeV,
$m_{1/2}=180$ GeV,
 $\xi_1=0.4$ (rad), $\xi_2=0.5$ (rad), $\xi_3=0.6$ (rad), $\alpha_{A_0}=0.8$ (rad) and $|A_0|=250$ GeV.}
\label{gamma3}
\end{figure}

\begin{figure}[t]
\hspace*{-0.8in}
\centering
\includegraphics[width=18cm,height=18cm]{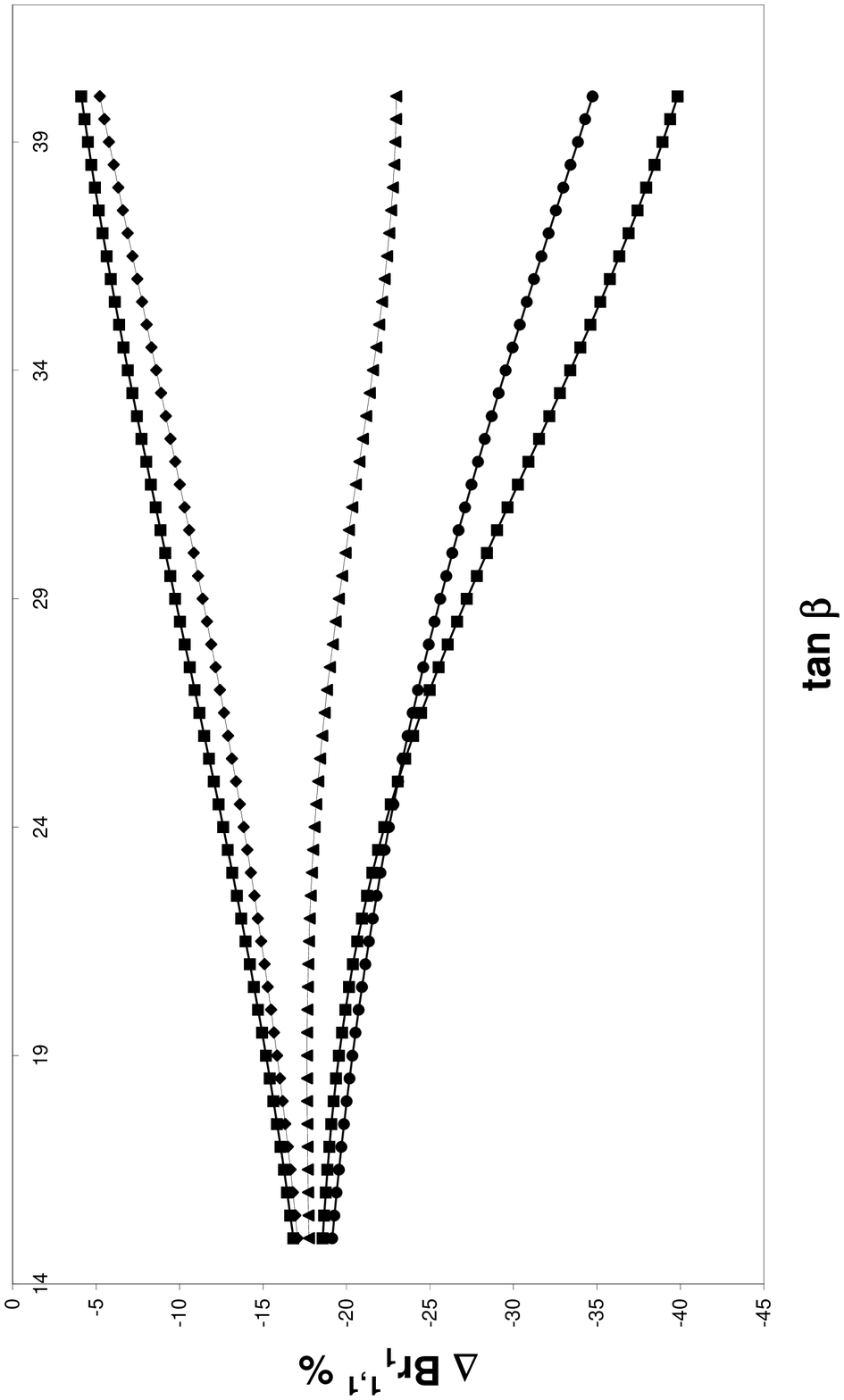}
\caption{$\tan\beta$ dependence of $\Delta Br_1\to \chi_1^+\chi_1^-$.
The curves in ascending order at $\tan\beta=40$ correspond to $\theta_{\mu}=0.5$, $0.1$, $1.0$, $1.5$ and $2.0$ (rad).
 The input is $m_0=500$ GeV,
$m_{1/2}=150$ GeV,
 $\xi_1=0.4$ (rad), $\xi_2=0.5$ (rad), $\xi_3=0.6$ (rad), $\alpha_{A_0}=0.3$ (rad) and $|A_0|=250$ GeV.}
\label{epsfig3a}
\end{figure}

\begin{figure}[t]
\hspace*{-0.8in}
\centering
\includegraphics[width=18cm,height=18cm]{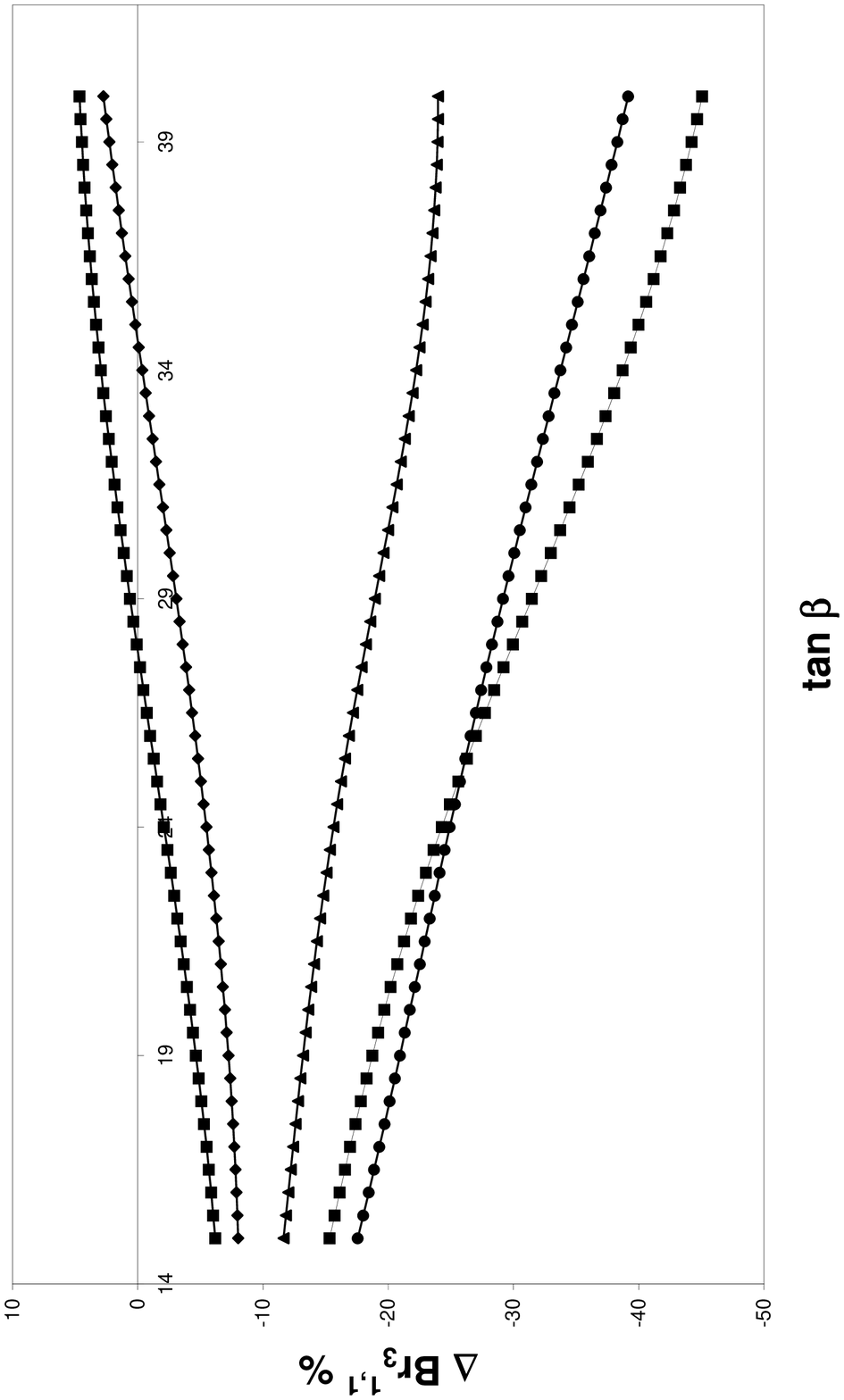}
\caption{$\tan\beta$ dependence of $\Delta Br_3\to \chi_1^+\chi_1^-$.
The curves in ascending order at $\tan\beta=40$ correspond to $\theta_{\mu}=0.5$, $0.1$, $1.0$, $1.5$ and $2.0 $(rad).
 The input is $m_0=500$ GeV,
$m_{1/2}=150$ GeV,
 $\xi_1=0.4$ (rad), $\xi_2=0.5$ (rad), $\xi_3=0.6$ (rad), $\alpha_{A_0}=0.3$ (rad) and $|A_0|=250$ GeV.}
\label{epsfig3c}
\end{figure}

\begin{figure}[t]
\hspace*{-0.8in}
\centering
\includegraphics[width=18cm,height=18cm]{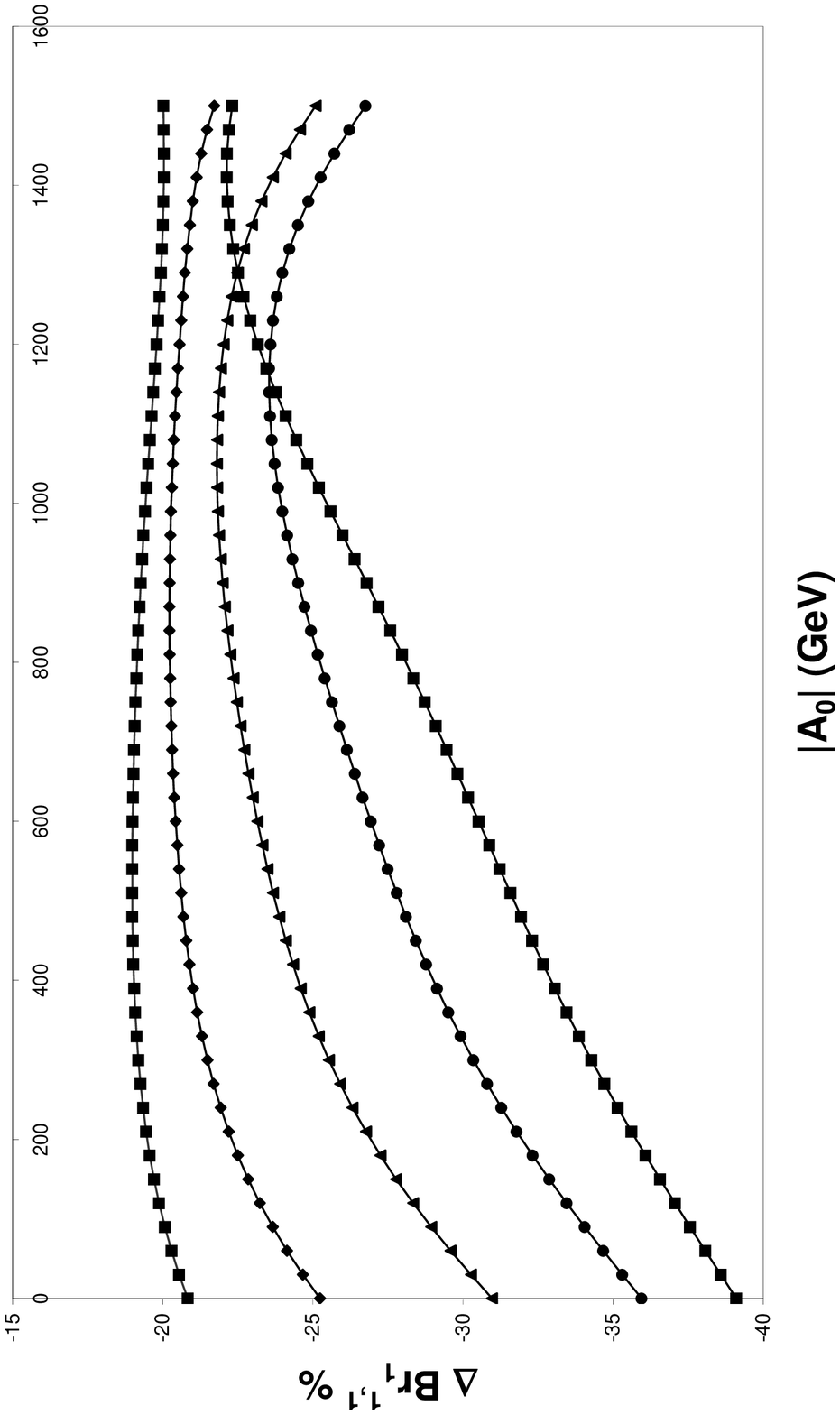}
\caption{$|A_0|$ dependence of $\Delta Br_1\to \chi_1^+\chi_1^-$.
The curves in ascending order  
at $|A_0|=0$ correspond
to $\tan\beta=40$, $35$, $30$, $25$ and $20$. The input is 
$m_0=500$ GeV,
$m_{1/2}=150$ GeV,
 $\xi_1=0.4$ (rad), $\xi_2=0.5$ (rad), $\xi_3=0.6$ (rad), $\theta_{\mu}=0.7$ (rad) and $\alpha_{A_0}=0.1$ (rad).}
\label{epsfig4a}
\end{figure}

\begin{figure}[t]
\hspace*{-0.8in}
\centering
\includegraphics[width=18cm,height=18cm]{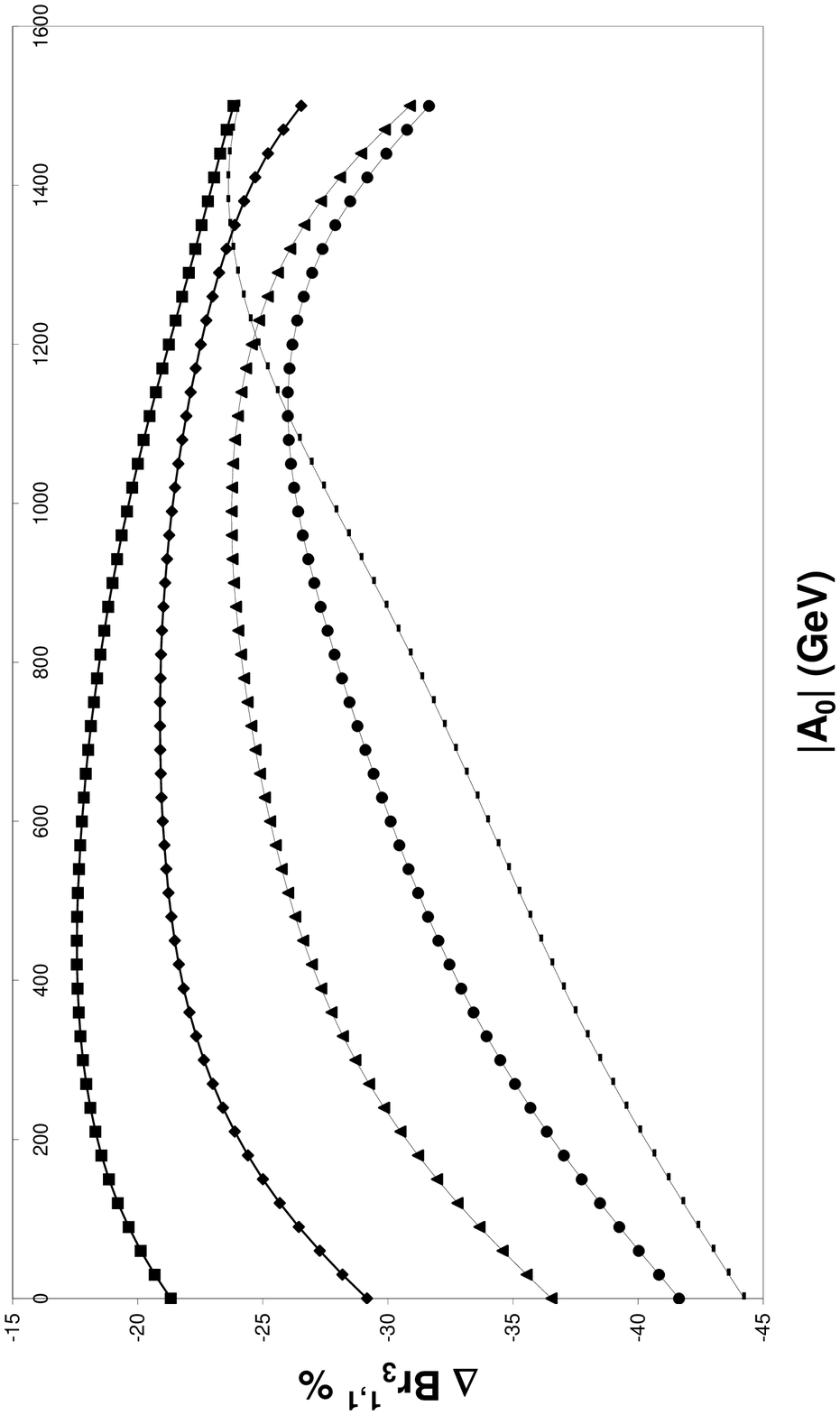}
\caption{$|A_0|$ dependence of $\Delta Br_3\to \chi_1^+\chi_1^-$.
The curves in ascending order  
at $|A_0|=0$ correspond
to $\tan\beta=40$, $35$, $30$, $25$ and $20$. The input is 
$m_0=500$ GeV,
$m_{1/2}=150$ GeV,
 $\xi_1=0.4$ (rad), $\xi_2=0.5$ (rad), $\xi_3=0.6$ (rad), $\theta_{\mu}=0.7$ (rad) and $\alpha_{A_0}=0.1$ (rad).}
\label{epsfig4c}
\end{figure}

\begin{figure}[t]
\hspace*{-0.8in}
\centering
\includegraphics[width=18cm,height=18cm]{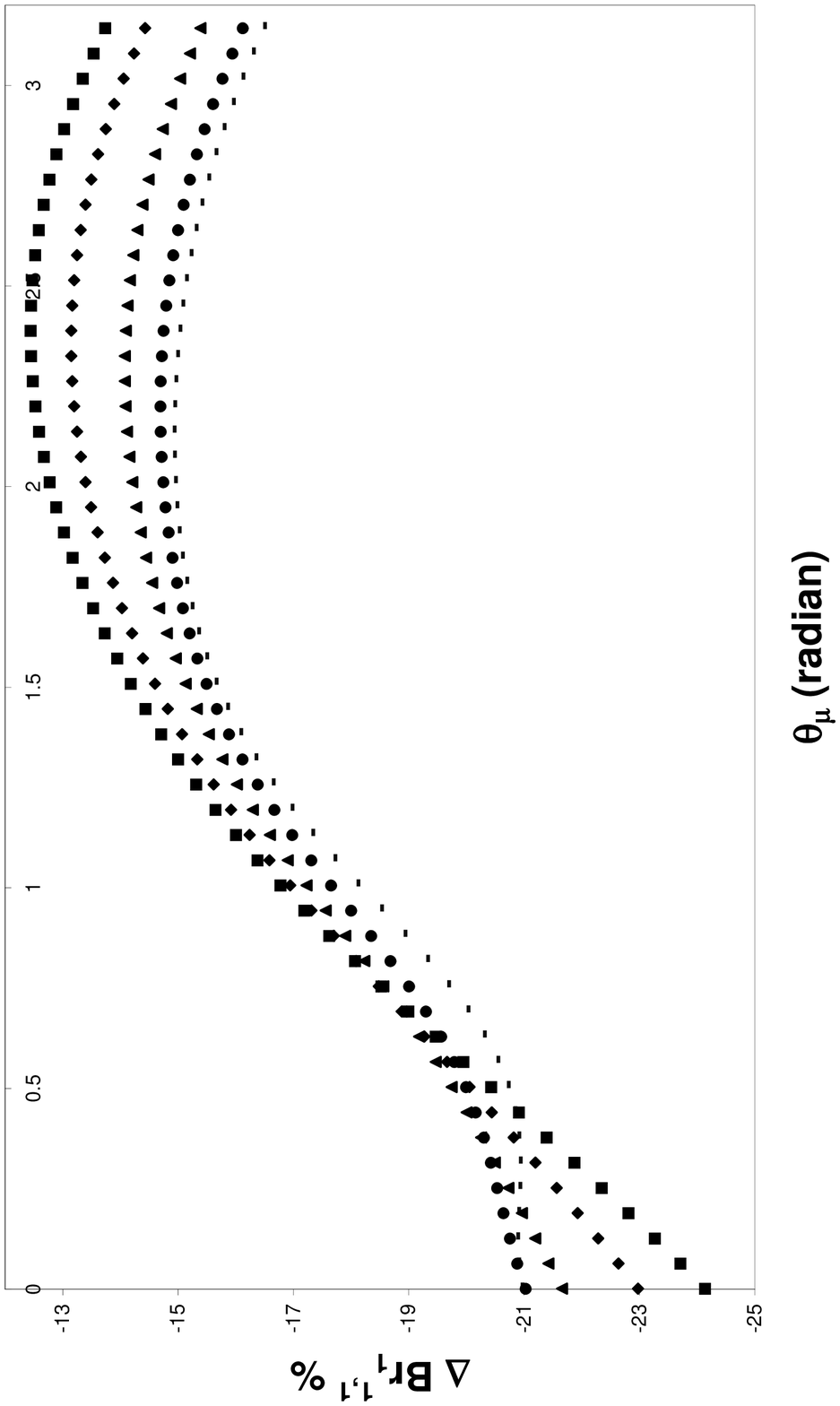}
\caption{$\theta_{\mu}$ dependence of $\Delta Br_1\to \chi_1^+\chi_1^-$.
The curves in ascending order
at $\theta_{\mu}=2.0$ (rad)
 correspond to $|A_0|=100$, $250$, $500$, $750$ and $900$ GeV. The input is
$\tan\beta=20.0$,
 $m_0=500$ GeV, 
$m_{1/2}=150$ GeV,
$\xi_1=0.4$ (rad), $\xi_2=0.5$ (rad), $\xi_3=0.6$ (rad) and $\alpha_{A_0}=0.2$ (rad).}
\label{epsfig5a}
\end{figure}

\begin{figure}[t]
\hspace*{-0.8in}
\centering
\includegraphics[width=18cm,height=18cm]{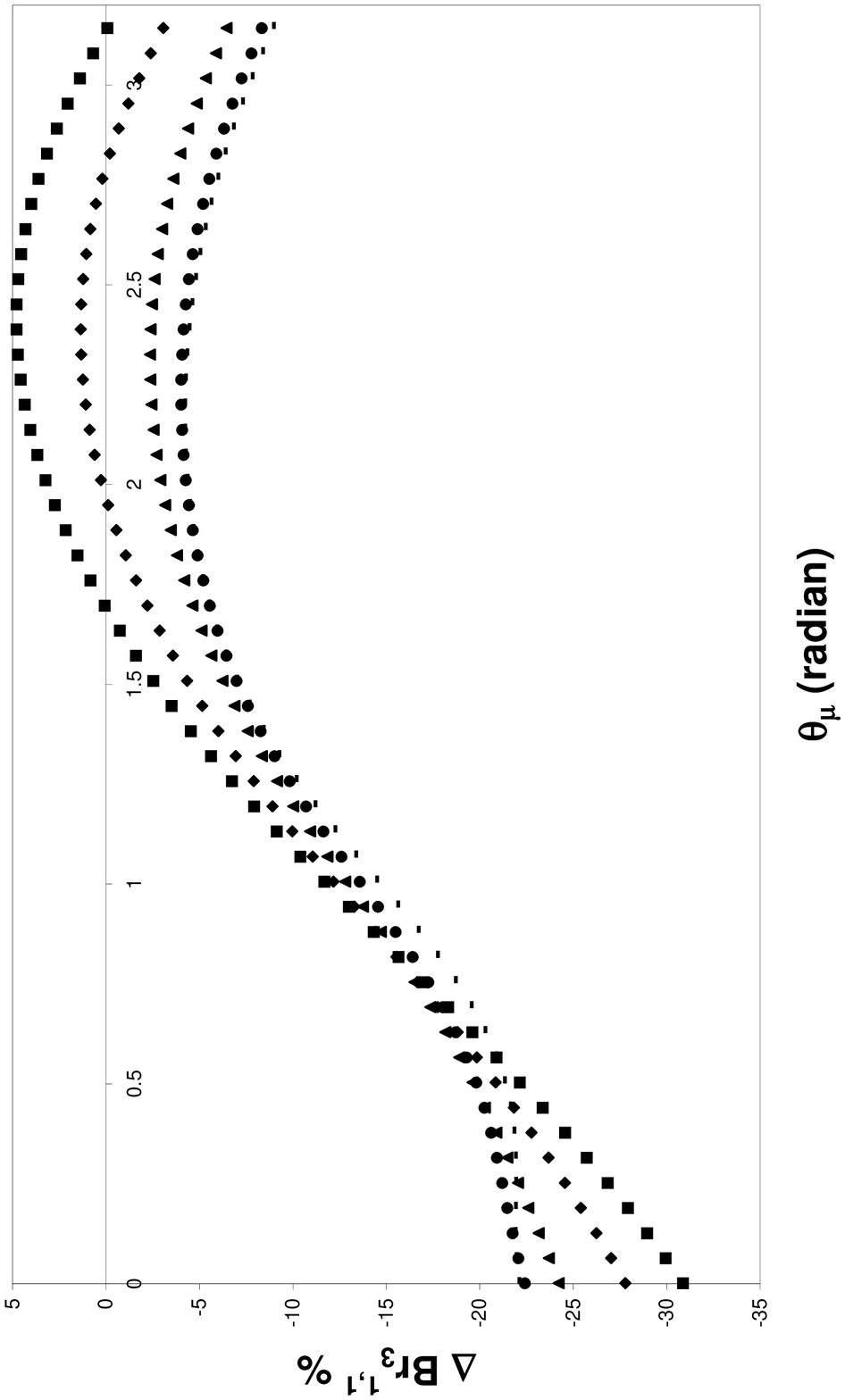}
\caption{$\theta_{\mu}$ dependence of $\Delta Br_3\to \chi_1^+\chi_1^-$.
The curves in ascending order
at $\theta_{\mu}=\pi$ (rad)
 correspond to $|A_0|=100$, $250$, $500$, $750$ and $900$ GeV. The input is
$\tan\beta=20.0$,
 $m_0=500$ GeV,
$m_{1/2}=150$ GeV,
 $\xi_1=0.4$ (rad), $\xi_2=0.5$ (rad), $\xi_3=0.6$ (rad) and $\alpha_{A_0}=0.2$ (rad).}
\label{epsfig5c}
\end{figure}

\begin{figure}[t]
\hspace*{-0.8in}
\centering
\includegraphics[width=18cm,height=18cm]{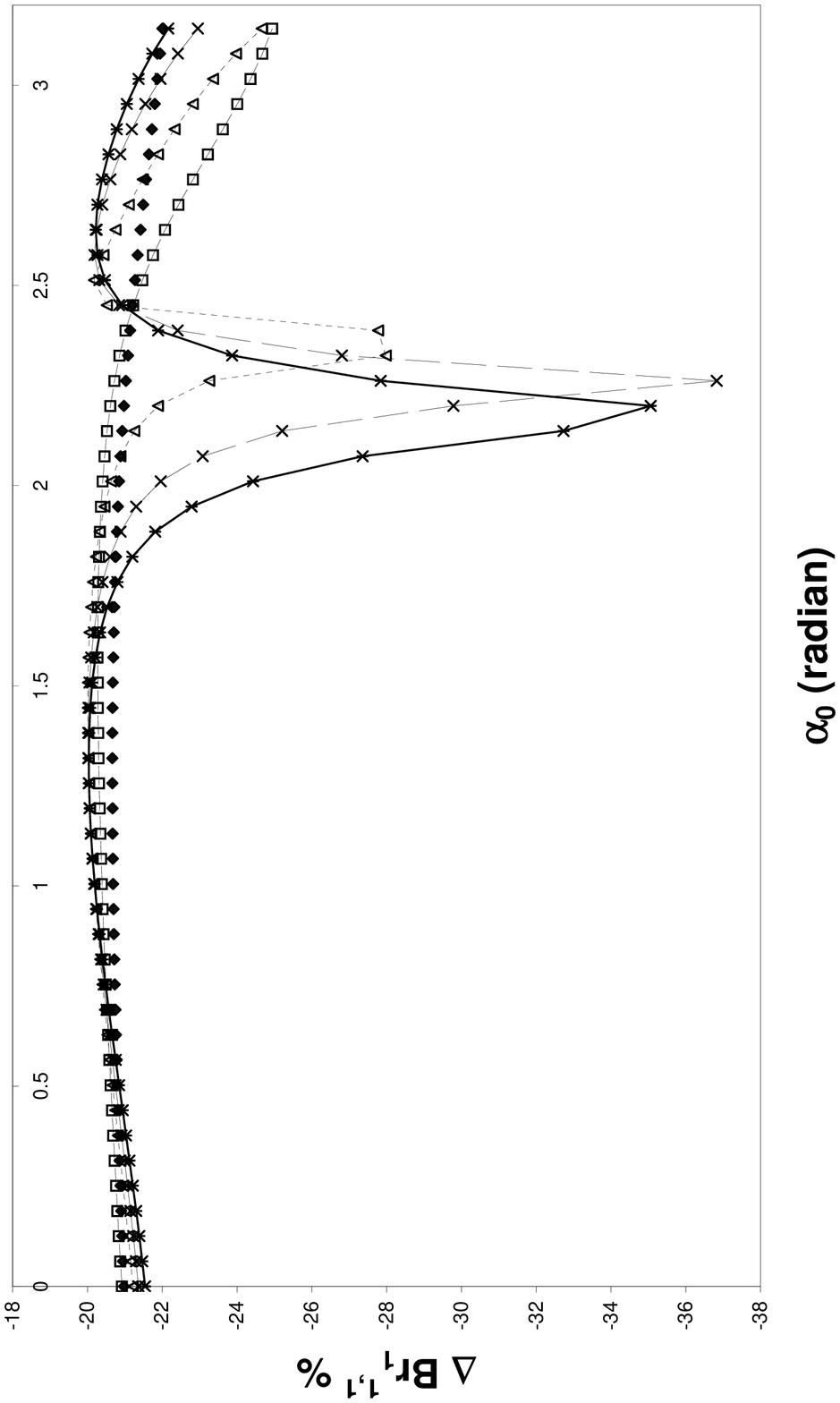}
\caption{$\alpha_0$ dependence of $\Delta Br_1\to \chi_1^+\chi_1^-$.
The curves in ascending order 
at $\alpha_{A_0}=2.2$ (rad)
correspond to $|A_0|=500$, $450$, $400$, $100$ and $200$ GeV. The input is
$\tan\beta=20.0$,
 $m_0=500$ GeV,
$m_{1/2}=150$ GeV,
 $\xi_1=0.4$ (rad), $\xi_2=0.5$ (rad), $\xi_3=0.6$ (rad) and $\theta_{\mu}=0.1$ (rad).}
\label{epsfig6a}
\end{figure}

\begin{figure}[t]
\hspace*{-0.8in}
\centering
\includegraphics[width=18cm,height=18cm]{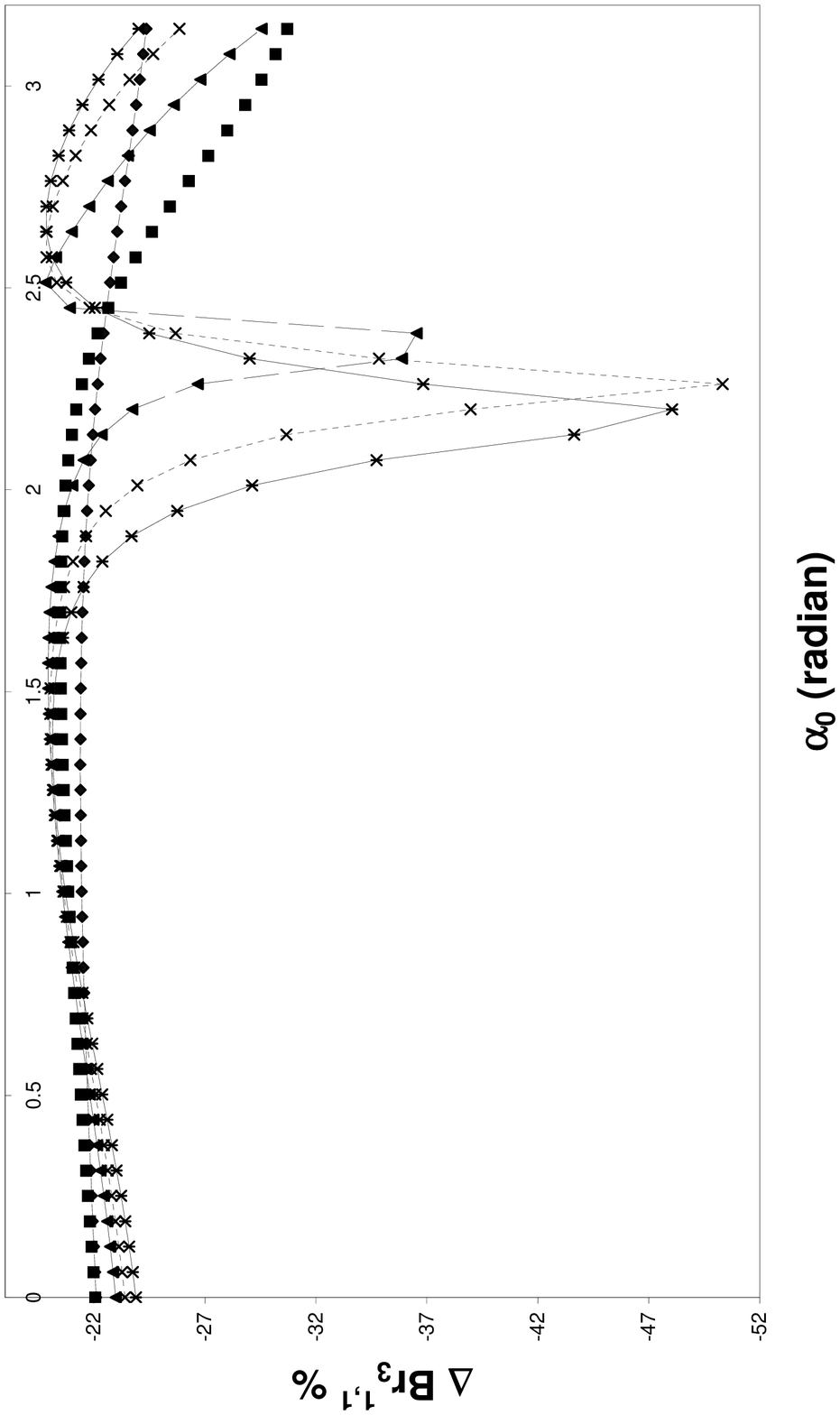}
\caption{$\alpha_0$ dependence of $\Delta Br_3\to \chi_1^+\chi_1^-$.
The curves in ascending order 
at $\alpha_{A_0}=2.2$ (rad)
correspond to $|A_0|=500$, $450$, $400$, $100$ and $200$ GeV. The input is
$\tan\beta=20.0$,
 $m_0=500$ GeV,
$m_{1/2}=150$ GeV,
 $\xi_1=0.4$ (rad), $\xi_2=0.5$ (rad), $\xi_3=0.6$ (rad) and $\theta_{\mu}=0.1$ (rad).}
\label{epsfig6c}
\end{figure}

\begin{figure}[t]
\hspace*{-0.8in}
\centering
\includegraphics[width=18cm,height=18cm]{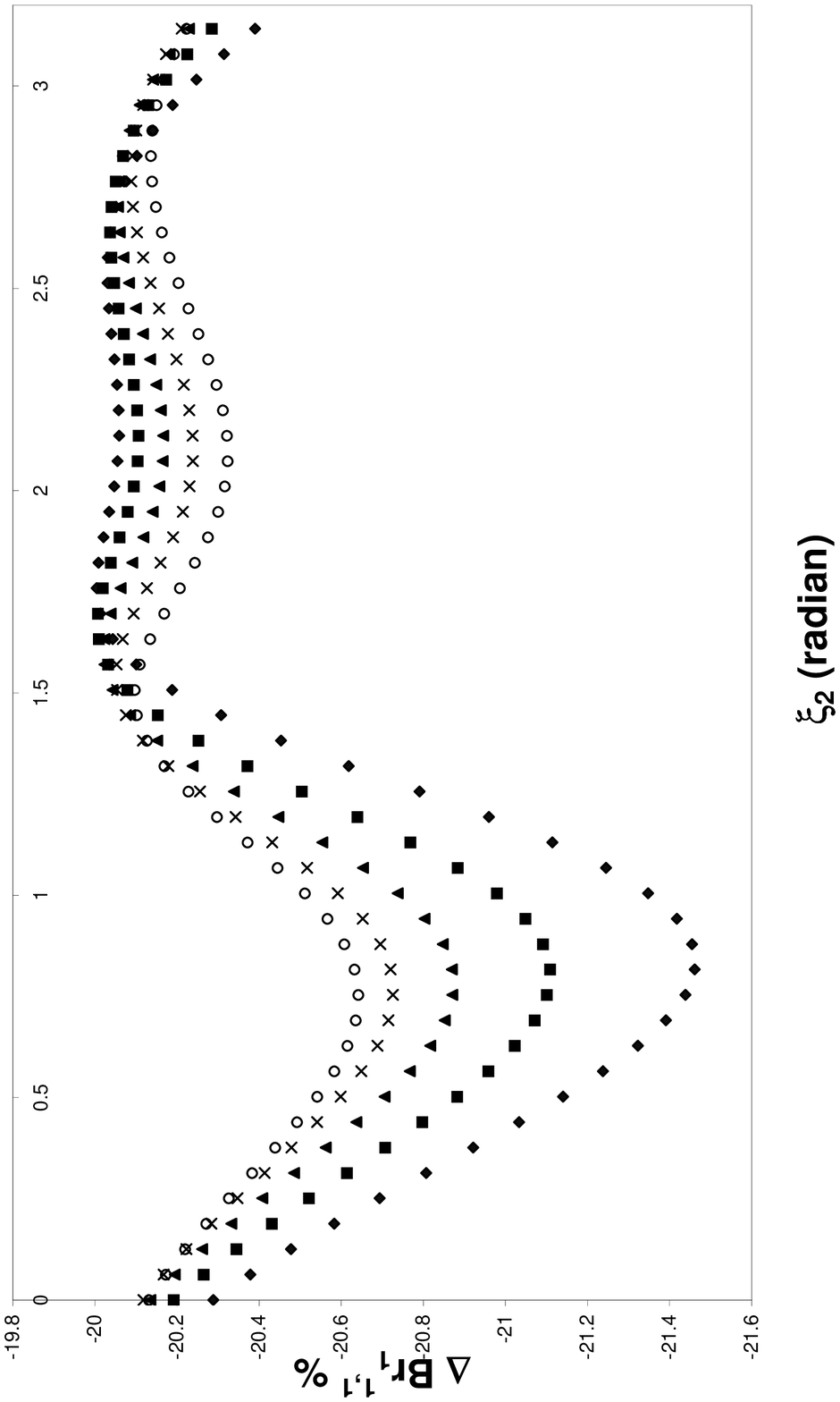}
\caption{$\xi_2$ dependence of $\Delta Br_1\to \chi_1^+\chi_1^-$.
The curves in ascending order 
at $\xi_2=0.75$ (rad)
correspond to $|A_0|=50$, $100$, $150$, $200$ and $250$ GeV. The input is
$\tan\beta=20.0$,
 $m_0=500$ GeV,
$m_{1/2}=150$ GeV,
 $\xi_1=0.4$ (rad), $\xi_3=0.6$ (rad) and $\theta_{\mu}=0.2$ (rad) and $\alpha_{A_0}=0.3$ (rad).}
\label{epsfig7a}
\end{figure}

\begin{figure}[t]
\hspace*{-0.8in}
\centering
\includegraphics[width=18cm,height=18cm]{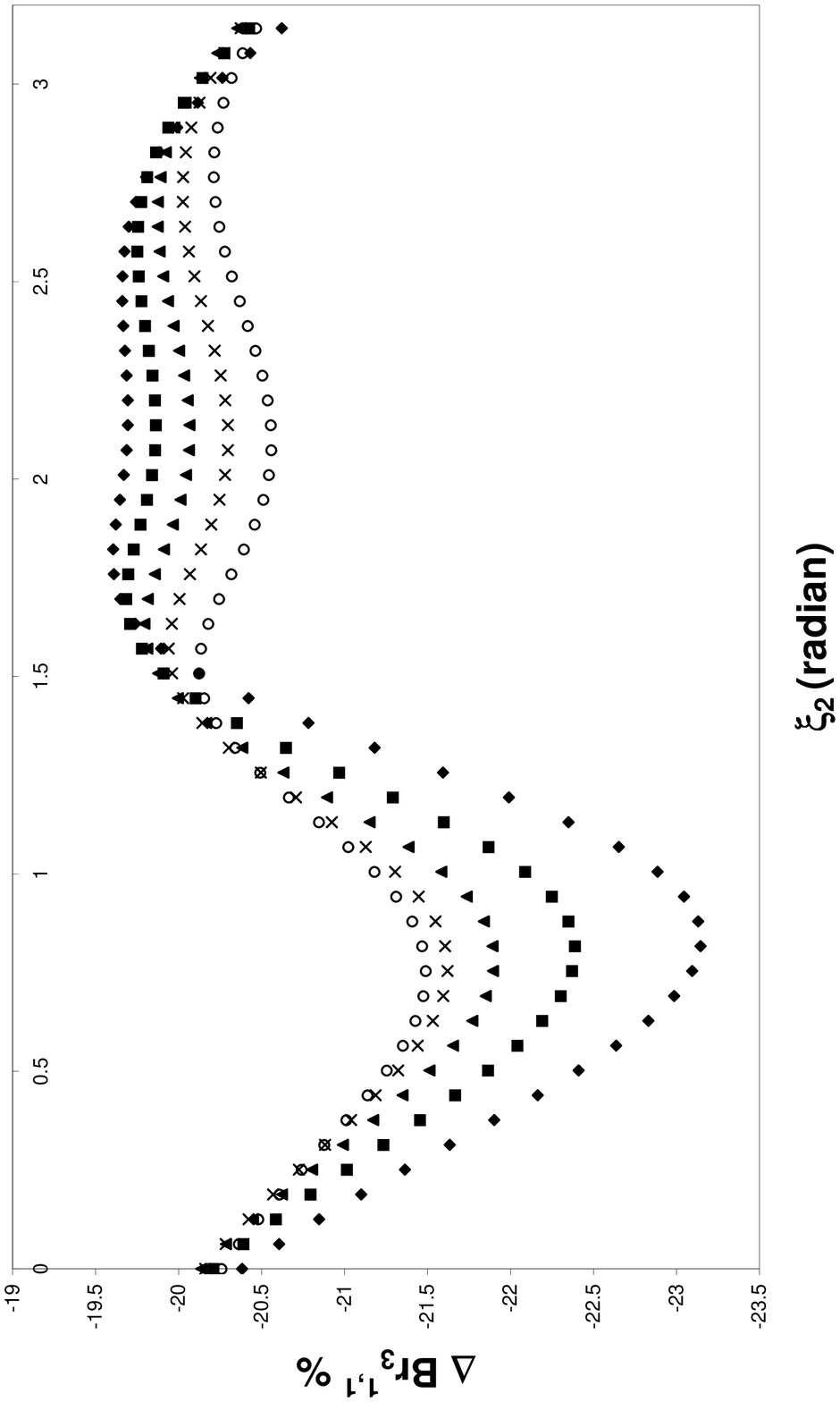}
\caption{$\xi_2$ dependence of $\Delta Br_3\to \chi_1^+\chi_1^-$.
The curves in ascending order 
at $\xi_2=0.75$ (rad)
correspond to $|A_0|=50$, $100$, $150$, $200$ and $250$ GeV. The input is
$\tan\beta=20.0$,
 $m_0=500$ GeV, 
$m_{1/2}=150$ GeV,
$\xi_1=0.4$ (rad), $\xi_3=0.6$ (rad) and $\theta_{\mu}=0.2$ (rad) and $\alpha_{A_0}=0.3$ (rad).}
\label{epsfig7c}
\end{figure}


\begin{thebibliography}{999}

\bibitem{carena1}
M. Carena and H. e. Haber, Prog. Part. Nucl. Phys. {\bf{50}}, 63 (2003).

\bibitem{edm1}
P. Nath, Phys. Rev. Lett.{\bf 66}, 2565(1991); 
Y. Kizukuri and  N. Oshimo, Phys.Rev.{\bf D46},3025(1992).
T. Ibrahim and P. Nath,  Phys.\ Lett.\ B {\bf 418}, 98 (1998); 
 Phys. Rev. {\bf D57}, 478(1998); Phys. Rev. {\bf D58}, 111301(1998);
 T. Falk and K Olive, Phys. Lett. {\bf B 439}, 71(1998);
 M. Brhlik, G.J. Good, and G.L. Kane, Phys. Rev. {\bf D59}, 115004
 (1999); A. Bartl, T. Gajdosik, W. Porod, P. Stockinger, and
 H. Stremnitzer,  Phys. Rev. {\bf 60}, 073003(1999);
 S. Pokorski, J. Rosiek and C.A. Savoy, 
 Nucl.Phys. {\bf B570}, 81(2000);
 E.~Accomando, R.~Arnowitt and B.~Dutta,
Phys.\ Rev.\ D {\bf 61}, 115003 (2000);
  U. Chattopadhyay, T. Ibrahim, D.P. Roy, Phys.Rev.D64:013004,2001.

\bibitem{edma}
 C.~S.~Huang and W.~Liao,
Phys.\ Rev.\ D {\bf 61}, 116002 (2000);
ibid, Phys.\ Rev.\ D {\bf 62}, 016008 (2000);
 A.Bartl, T. Gajdosik, E.Lunghi, A. Masiero, W. Porod,
H. Stremnitzer and O. Vives, hep-ph/0103324.
 M. Brhlik, L. Everett, G. Kane and J. Lykken, Phys. Rev.
 Lett. {\bf 83}, 2124, 1999; Phys. Rev. {\bf D62}, 035005(2000);
  E. Accomando, R. Arnowitt and B. Dutta, 
Phys. Rev. {\bf D61},  075010(2000);
T. Ibrahim and P. Nath, Phys. Rev. {\bf D61}, 093004(2000).

\bibitem{edmb}
T. Falk, K.A. Olive, M. Prospelov, and R. Roiban, Nucl. Phys. 
 {\bf B560}, 3(1999); V.~D.~Barger, T.~Falk, T.~Han, J.~Jiang, T.~Li 
 and T.~Plehn,
Phys.\ Rev.\ D {\bf 64}, 056007 (2001);
S.Abel, S. Khalil, O.Lebedev, Phys. Rev. Lett. {\bf 86}, 5850(2001);
T.~Ibrahim and P.~Nath,
Phys.\ Rev.\ D {\bf 67}, 016005 (2003)
arXiv:hep-ph/0208142.
D. Chang, W-Y.Keung,and A. Pilaftsis, Phys. Rev. Lett. {\bf 82}, 
900(1999). 
\bibitem{edm2}
E. Commins et al., Phys. Rev. A {\bf 50}, 2960(1994)
\bibitem{edm3}
P. G. Harris et al., Phys. Rev. Lett. {\bf 82}, 904(1999)
\bibitem{edm4}
S. K. Lamoreaux, J. P. Jacobs, B. R. Heckel, F. J. Raab, and E. N. Forston, Phys. Rev. Lett. {\bf 57}, 3125(1986).
\bibitem{edm5}
A. Pilaftsis, Phys. Rev. {\bf D58}, 096010; Phys. Lett.{\bf B435}, 
88(1998);
~A. Pilaftsis and C.E.M. Wagner, Nucl. Phys. {\bf B553}, 3(1999);
~D.A. Demir, Phys. Rev. {\bf D60}, 055006(1999);
~S.~Y.~Choi, M.~Drees and J.~S.~Lee,
Phys.\ Lett.\ B {\bf 481}, 57 (2000);
T.~Ibrahim,
Phys.\ Rev.\ D {\bf 64}, 035009 (2001).

\bibitem{edm5a}
~S.~W.~Ham, S.~K.~Oh, E.~J.~Yoo, C.~M.~Kim and D.~Son,
arXiv:hep-ph/0205244;
~M.~Boz,
Mod.\ Phys.\ Lett.\ A {\bf 17}, 215 (2002).

\bibitem{edm5b}
M.~Carena, J.~R.~Ellis, A.~Pilaftsis and C.~E.~Wagner,
Nucl.\ Phys.\ B {\bf 625}, 345 (2002)
[arXiv:hep-ph/0111245].
J.~Ellis, J.~S.~Lee and A.~Pilaftsis,
arXiv:hep-ph/0404167.
E. Chrisova, H. Eberl, W. Majerotto, and S. Kraml, J. High Energy Phys. 12 (2002)021; E. Christova, H. Eberl, W. Majerotto, and S. Kraml, Nucl. Phys. B {\bf 639}, 263(2002); {\bf 647}, 359(E) (2002)
T. Ibrahim, P. Nath, Phys and A. Psinas. Rev. D {\bf 70}, 035006(2004).
\bibitem{edm6}
U. Chattopadhyay, T. Ibrahim and P. Nath, 
 Phys. Rev. {\bf D60},063505(1999); 
 T. Falk, A. Ferstl and K. Olive, Astropart. Phys. {\bf 13}, 301(2000);
 S. Khalil, Phys. Lett. {\bf B484}, 98(2000);
S. Khalil and Q. Shafi, Nucl.Phys. {\bf B564}, 19(1999);
K. Freese and P. Gondolo, hep-ph/9908390;
 S.Y. Choi, hep-ph/9908397;
M.~E.~Gomez, T.~Ibrahim, P.~Nath and S.~Skadhauge,
Phys. Rev. D {\bf 70}, 035014 (2004); T. Nihei and M. sasagawa, Phys. Rev. D {\bf 70}, 055011(2004); {\bf 70}, 079901(2004).

\bibitem{edm7}
 M. Gomez, T. Ibrahim, P. Nath and S. Skadhauge, Phys. Rev. D{\bf 74}, 015015(2006).

\bibitem{cpreview}
 T.~Ibrahim and P.~Nath,
  %``CP violation from standard model to strings,''
  hep-ph/0705.2008.
  %%CITATION = ARXIV:0705.2008;%%


\bibitem{sugra1}
A. H. Chamseddine, R. Arnowitt, and P. nath, Phys. Rev. Lett. {\bf 49}, 970(1982); 
R. Barbieri, S. Ferrara, and C. a. Savoy, Phys. Lett. B {\bf 119}, 343(1982); L. Hall,
J. Lykken, and S. weinberg, Phys. Rev. D {\bf 27}, 2359(1983); P. Nath, R. Arnowitt,
and A. H. Chamseddine, Nucl. Phys. B {\bf 227}, 121(1983).

\bibitem{babu}
K. S. Babu and C. F. Kolda, Phys. Lett. B{\bf 451}, 77, 1999.

\bibitem{we1}
T. Ibrahim, P. Nath, Phys. Rev. D {\bf 68}, 015008(2003).

\bibitem{two}
D. A. Demir and K. A. Olive, Phs. Rev. {\bf 65}, 034007 (2002);
G. Degrassi, P. Gambino, and G. F. Giudice, J. High Energy Phys. {\bf 12}, 009(2000);
G. Belanger, F. Boudjema, A. Pukhov, and A. Semenov, Comput. Phys. Commun. {\bf 149}, 103(2002).
\bibitem{we2}
T. Ibrahim, P. Nath, Phys. Rev. D {\bf 69}, 075001(2004)

\bibitem{we3}
T. Ibrahim, P. Nath and A. Psinas, Phys. Rev. D {\bf 70}, 
035006(2004).

\bibitem{eberl}
H. Eberl, W. Majerotto, Y. Yamada, Phys. Lett. B {\bf 597}
(2004) 275.

\bibitem{ren}
 Z. Ren-You, M. Wen-Gan, W. Lang-Hui and J. Yi, Phys. Rev. D {\bf 65}, 075018(2002).

\bibitem{demir1}
D. A. Demir, Phys. Rev. D {\bf 60}, 055006(1999).

\bibitem{we4}
T. Ibrahim, P. Nath, Phys. Rev. D {\bf 63}, 035009(2001).

\bibitem{we5}
T. Ibrahim, P. Nath, Phys. Rev. D {\bf 66}, 015005(2002).

\bibitem{gunion}
J. F. Gunion, H. E. Haber, Nucl. Phys. B {\bf 307}, 445 (1988);
A. Djouadi, hep/ph/9712334; A. Djouadi, J. Kalinowski, and M. Spira, 
Comp. Phys. Commun. {\bf 108}, 56 (1998).

\bibitem{we6}
The neutralino decay of neutral Higgs with CP  phases will be discussed elsewhere.

\bibitem{allan}
B. C. Allanach et al., Eur. Phys. J. C {\bf 25}, 113 (2002).

\bibitem{last1}
M. Carena, M. Olechowski, S. Pokorski and C. E. M. Wagner, Nucl. Phys. B{\bf 426} (1994) 269; L. J. Hall, R. Rattazzi and O. Sarid, Phys. Rev. D{\bf 50} (1994) 7048;
T. Ibrahim and P. Nath, Phys. Rev. D {\bf 67}, 095003(2003).


\end{thebibliography}
\end{document}